\documentclass[lettersize,journal]{IEEEtran}

\usepackage{ifpdf}
\usepackage{cite}
\usepackage[pdftex]{graphicx}

\usepackage[T1]{fontenc}
\usepackage{afterpage}
\usepackage{bm}
\usepackage{listings}
\usepackage{xcolor}
\usepackage{amsthm}
\usepackage{graphicx}
\usepackage{float}
\usepackage[font=small]{caption}
\usepackage[font=footnotesize]{subcaption}

\usepackage{placeins}
\usepackage{url}
\usepackage{epstopdf}
\usepackage{booktabs}
\usepackage{footnote}
\makesavenoteenv{table}
\makesavenoteenv{tabular}
\usepackage{flushend}
\usepackage{amsmath}
\usepackage{amssymb}
\usepackage{tabularx}
\usepackage{booktabs}
\usepackage{placeins}
\usepackage{graphicx}
\usepackage{color}
\usepackage{colortbl}
\usepackage{bbold}
\DeclareMathOperator*{\argmax}{arg\,max}
\usepackage[acronyms,nonumberlist,nopostdot,nomain,nogroupskip]{glossaries}
\usepackage{tabularx}
\usepackage{paralist}

\usepackage{tikz}
    \usetikzlibrary{shapes.arrows}

\usepackage{pgfplots}
\usepackage{adjustbox}
\usepackage{multirow}
\usepackage{multicol}

\usepackage{cuted}

\usepackage{gincltex}
\usepgfplotslibrary{fillbetween}
\usetikzlibrary{plotmarks}
\usetikzlibrary{patterns}
\usepackage{mathtools}
\usepackage{booktabs}
\usepackage{threeparttable}
\usepackage{amsfonts}
\DeclareMathSymbol{\shortminus}{\mathbin}{AMSa}{"39}

\usetikzlibrary{external}
\newlength{\hatchspread}
\newlength{\hatchthickness}
\newlength{\hatchshift}
\newcommand{\hatchcolor}{}
\tikzset{hatchspread/.code={\setlength{\hatchspread}{#1}},
         hatchthickness/.code={\setlength{\hatchthickness}{#1}},
         hatchshift/.code={\setlength{\hatchshift}{#1}},
         hatchcolor/.code={\renewcommand{\hatchcolor}{#1}}}
\tikzset{hatchspread=3pt,
         hatchthickness=0.8pt,
         hatchshift=0pt,
         hatchcolor={blue!20}}
\pgfdeclarepatternformonly[\hatchspread,\hatchthickness,\hatchshift,\hatchcolor]
   {custom north west lines}
   {\pgfqpoint{\dimexpr-2\hatchthickness}{\dimexpr-2\hatchthickness}}
   {\pgfqpoint{\dimexpr\hatchspread+2\hatchthickness}{\dimexpr\hatchspread+2\hatchthickness}}
   {\pgfqpoint{\dimexpr\hatchspread}{\dimexpr\hatchspread}}
   {
    \pgfsetlinewidth{\hatchthickness}
    \pgfpathmoveto{\pgfqpoint{0pt}{\dimexpr\hatchspread+\hatchshift}}
    \pgfpathlineto{\pgfqpoint{\dimexpr\hatchspread+0.15pt+\hatchshift}{-0.15pt}}
    \ifdim \hatchshift > 0pt
      \pgfpathmoveto{\pgfqpoint{0pt}{\hatchshift}}
      \pgfpathlineto{\pgfqpoint{\dimexpr0.15pt+\hatchshift}{-0.15pt}}
    \fi
    \pgfsetstrokecolor{\hatchcolor}
    \pgfusepath{stroke}
   }

\renewcommand{\epsilon}{\varepsilon}
\def \fwidth{\linewidth}

\def \fheight{0.55\linewidth}
\def \lfheight{0.55\linewidth}

\def \tfwidth{0.99\linewidth}
\def \tfheight{0.8\linewidth}

\definecolor{violet}{rgb}{0.6,0,0.6}%
\definecolor{orange_D}{rgb}{1,0.3,0}%
\definecolor{cyan}{rgb}{0,0.67,0.64}%
\definecolor{red}{rgb}{0.9,0,0}%
\definecolor{green}{rgb}{0,0.8,0}%
\definecolor{green_D}{rgb}{0,0.6,0}%
\definecolor{yellow}{rgb}{1,0.8,0}

\usepackage{glossaries}

\usepackage{cleveref}
\crefname{section}{Sec.}{sec.}
\crefname{figure}{Fig.}{fig.}
\crefname{equation}{Eq.}{eq.}

\newacronym{cbr}{CBR}{Constant Bit Rate}
\newacronym[\glslongpluralkey={Markov Decision Processes}]{mdp}{MDP}{Markov Decision Process}
\newacronym{pdf}{PDF}{Probability Density Function}
\newacronym{cdf}{CDF}{Cumulative Distribution Function}
\newacronym{mc}{MC}{Markov Chain}
\newacronym{edf}{EDF}{Earliest Deadline First}
\newacronym{hlf}{HLF}{Highest Level First}
\newacronym{qos}{QoS}{Quality of Service}
\newacronym{mptcp}{MPTCP}{Multi-path TCP}
\newacronym{lowrtt}{LowRTT}{lowest RTT first}
\newacronym{dems}{DEMS}{Decoupled Multipath Scheduler}
\newacronym{stms}{STMS}{Slide Together Multipath Scheduler}
\newacronym{daps}{DAPS}{Delay Aware Packet Scheduling}
\newacronym{blest}{BLEST}{Blocking Estimation}
\newacronym{iid}{IID}{Independent and Identically Distributed}
\newacronym{rtt}{RTT}{Round-Trip Time}
\newacronym{fec}{FEC}{Forward Error Correction}
\newacronym{leap}{LEAP}{Latency-controlled End-to-End Aggregation Protocol}
\newacronym{urllc}{URLLC}{Ultra-Reliable Low Latency Communications}
\newacronym{ccr}{CCR}{Constant Coding Rate}
\newacronym{ps}{PS}{Plain Split}
\newacronym{pec}{PEC}{Packet Erasure Channel}
\newacronym{pmf}{PMF}{Probability Mass Function}
\newacronym{hop}{HOP}{High-reliability latency-bounded Overlay Protocol}
\newacronym{vr}{VR}{Virtual Reality}
\newacronym{qs}{QS}{Queuing System}
\newacronym{iot}{IoT}{Internet of Things}
\newacronym{v2x}{V2X}{Vehicle to Everything}

\pgfplotsset{compat=1.15}

\begin{document}

\title{Optimal Latency-Oriented Coding and Scheduling in Parallel Queuing Systems}

\author{Andrea Bedin,
        Federico Chiariotti,~\IEEEmembership{Member,~IEEE,}
        Stepan Kucera,~\IEEEmembership{Senior Member,~IEEE,}
        and Andrea Zanella,~\IEEEmembership{Senior Member,~IEEE}
\thanks{Andrea Bedin (corresponding author, andrea.bedin.2@phd.unipd.it) and Andrea Zanella (andrea.zanella@unipd.it) are with the Dept. of Information Engineering, University of Padova, Italy. Federico Chiariotti (fchi@es.aau.dk) is with the Dept. of Electronic Systems, Aalborg University, Denmark. Stepan Kucera is with Nokia Bell Labs, Munich, Germany (stepan.kucera@nokia.com). Andrea Bedin is also with Nokia Bell Labs, Espoo, Finland. This work has received funding from the European Union's EU Framework Programme for Research and Innovation Horizon 2020 under Grant Agreement No 861222.}
}
\maketitle

\begin{abstract}
The evolution of 5G and Beyond networks has enabled new applications with stringent end-to-end latency requirements, but providing reliable low-latency service with high throughput over public wireless networks is still a significant challenge. One of the possible ways to solve this is to exploit path diversity, encoding the information flow over multiple streams across parallel links. The challenge presented by this approach is the design of joint coding and scheduling algorithms that adapt to the state of links to take full advantage of path diversity.
In this paper, we address this problem for a synchronous traffic source that generates data blocks at regular time intervals (e.g., a video with constant frame rate) and needs to deliver each block within a predetermined deadline. We first develop a closed-form performance analysis in the simple case of two parallel servers without any buffering and single-packet blocks, and propose a model for the general problem based on a \gls{mdp}. We apply policy iteration to obtain the coding and scheduling policy that maximizes the fraction of source blocks delivered within the deadline: our simulations show the drawbacks of different commonly applied heuristic solutions, drawing general design insights on the optimal policy.
\end{abstract}

\glsresetall

\section{Introduction}

\begin{figure*}[!t]
 \centering
          \input{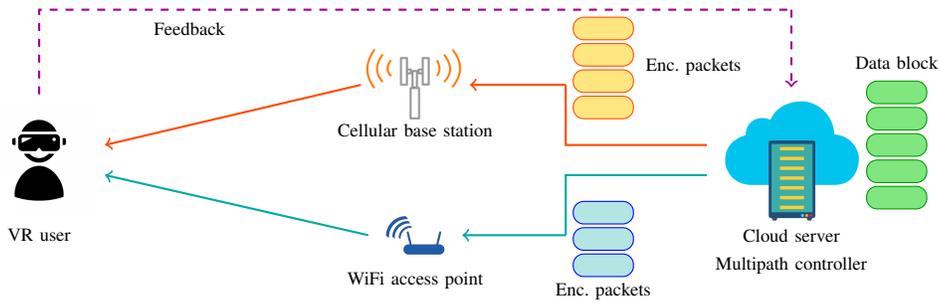}
 \caption{Schematic of an encoded multipath transmission: the 5 green packets are encoded into 7 and divided between the cellular link (4 yellow packets) and the WiFi access point (3 blue packets). The user will receive the data block as soon as any 5 packets are transmitted successfully.}
 \label{fig:multipath_scenario}
\end{figure*}

Over the past few years, the evolution of 5G and Beyond networks has opened new possibilities for interactive applications, such as mobile \gls{vr} or remote control of industry machinery, which were previously constrained to wired scenarios. Besides a fairly large transmission capacity, these applications have strict latency constraints which may be difficult to meet (e.g., the interactivity requirement for \gls{vr} or video conferencing) over wireless links because of the volatile nature of the medium, with fluctuating capacity and a relatively high packet error probability. The use of multiple wireless interfaces, often over different technologies, is a way to provide the required \gls{qos} even when individual links are unreliable. Indeed, encoding data blocks and sending redundant information over multiple paths can protect the transmission from failures and delays on individual paths. 
A possible example of reliable multipath communication is depicted in Fig.~\ref{fig:multipath_scenario}, in which a \gls{vr} user receives a stream of frames from a remote server with strict real-time requirements. The primary 5G link might not be sufficient, particularly in underserved rural areas, and 4G or private WiFi can be used to provide additional reliability. The end-to-end connections between the client and server can be affected by several factors, such as propagation and mobility issues or cross traffic. The objective of this paper is to provide a theoretical model of such a scenario, abstracting each link to a queuing model to find the optimal schedule to reliably transmit the data with bounded latency. The existing \gls{mptcp} standard is woefully inadequate for this reason: improper scheduling can cause significant delays due to the head of line blocking problem, and retransmissions can compound the problem, often providing worse \gls{qos} than even a single-path flow on the best available path~\cite{ferlin2014multi}. The practical applications of our work are in real-time solutions that avoid retransmissions, relying on packet-level coding to protect the transmission.

While efficient ways to exploit multiple paths to reliably transmit \gls{urllc} traffic~\cite{nielsen2017ultra} exist, they are limited to applications with very low throughput and very tight delay constraints, while applications with looser real-time constraints, but a far higher data rate, mostly operate on a best-effort basis.
In our work, we focus on this type of sources, considering a heavy flow with periodic block arrivals and a tight latency constraint, such as \gls{vr} streams or sensor data flows generated by, e.g., autonomous vehicles~\cite{chiariotti2021quic} or high-throughput industrial closed-loop control data~\cite{TS22104}.
Recently, there has been an effort to provide reliable end-to-end service using redundant coding over multiple paths: in particular, the \gls{dems} \gls{mptcp} scheduler~\cite{guo2017accelerating} introduced the notion of data blocks, which must be delivered as a whole, and exploited packet-level coding across different paths to ensure a faster, more reliable delivery of the data, recovering from failures on any single path, and our previous work~\cite{chiariotti2019analysis} introduces a dynamic coding rate adaptation to the available capacity. However, even state-of-the-art protocols still use \emph{ad hoc} heuristic mechanisms to balance the tradeoff between maintaining a high reliability (which would require more redundancy) and avoiding self-queuing delays for future blocks (which inherently limits the possible redundancy that can be sent over the available paths). In this context, the tradeoff is extremely complex: adding too much redundancy on the wrong path can cause congestion (if the capacity of the path is exceeded), while adding too little can reduce reliability and make the transmission less robust to errors or capacity fluctuations. The interaction between queues in a multipath system is even more complex, and optimizing decisions even in a relatively idealized scenario is a formidable problem. To the best of our knowledge, this is the first work in the literature to rigorously model this tradeoff and optimize coding and scheduling jointly, considering a simplified scenario that can provide general insights for more practical future schemes. Existing theoretical models~\cite{squillante2007stochastic} often use static scheduling policies and attempt to derive bounds on the latency~\cite{joshi2015queues}, but a joint optimization of coding and scheduling has never been attempted.

In this paper, we model a multipath communication scenario as a fork-join \gls{qs}, i.e., a system in which packets from the same flow are sent over multiple queues and gathered by the receiver. We then cast the queuing model into a \gls{mdp} framework to capture the long-term effects of the controller's decisions. We solve the joint coding and scheduling \gls{mdp} to derive the optimal policy, deciding \emph{how much} redundancy to add to each data block and \emph{how to split} the coded data among the available paths. In other words,  the controller has a dual objective: firstly, it needs to determine the necessary amount of redundancy for the data block, and secondly, it needs to act as an optimal scheduler, distributing the encoded packets over a number of parallel queues in order to deliver the data within a fixed deadline with stochastic reliability guarantees, i.e., to maintain below a certain threshold the probability that the data block latency exceeds the deadline. We also consider the implications of packet loss and delayed feedback on the state of the queue, as well as time-varying queues that can model different wireless scenarios.
In fact, one of the main potential drawbacks of redundancy is to build-up packet queues at temporarily slower links. This might trigger a ``snowball'' effect, as schedulers try to react to the increase in the delay due to the queued packets by adding even more redundancy, as observed practically in~\cite{chiariotti2019analysis}. Therefore, the optimal policy must properly account for self-inflicted latency, striking a balance between immediate and long-term reward.

The main contribution of our work is a rigorous mathematical model of the parallel communication as a fork-join queuing problem~\cite{kim1989fork}, with a full derivation of the reliability and latency violation probability. We find the optimal solution to the problem explicitly in a simple case and by policy iteration in the more general scenario. While policy iteration is computationally heavy, it is provably optimal and analytically tractable, and practical solutions can use more flexible reinforcement learning methods. The optimality gap in our model can give important insights on the design of practical algorithm, and while the model's complexity is limited by the tractability of the equations, its features can be used to simulate a wide array of potential scenarios, including cases with delayed feedback or incorrect channel parameter estimation. Since, to the best of our knowledge, the literature is currently lacking theoretical tools to analyze these scenarios, our analysis can spur further development in coded multipath communications.

The rest of the paper is structured as follows: first, we examine the state of the art on parallel \glspl{qs} while their practical applications are discussed in Sec.~\ref{sec:related}. We then describe the generic system model for the considered system in \cref{sec:sysmodel}. \cref{sec:mdp} defines the \gls{mdp} formulation of the problem and presents its solution in the most general case. We also present the result for some notable reward functions, such as the overall expected amount of data delivered within the deadline. Then, Sec.~\ref{sec:analytical} presents the derivation of an analytical policy for a simple example in which it is possible to do so. In \cref{sec:results}, we show our simulation results, including an analysis of the optimal policy and its comparison against some practical heuristics taken from the literature, such as load balancing and max redundancy. Finally, \cref{sec:conc} concludes the paper.

\section{Related Work}\label{sec:related}

The stochastic characterization of parallel queues with multiple servers~\cite{grassmann1980transient} has gradually become an active research subject~\cite{squillante2007stochastic}, following the development of parallel computing and multipath networking. The problem of scheduling Poisson arrivals from multiple sources on multiple queues, minimizing the response time for each source, can be solved using classical nonlinear optimization~\cite{sethuraman1999optimal}. Policies can be found even if the state of the queues is not directly observable, getting a maximum likelihood estimate from the known capacity distributions~\cite{konovalov2017using} or employing periodic policies~\cite{anselmi2015control}.

As mentioned above, our work deals with the \emph{fork-join} queuing model~\cite{kim1989fork}, where incoming tasks or blocks of data are divided among several parallel \glspl{qs} with independent queues~\cite{khudabukhsh2017optimizing}. The first work to find the latency bounds for the transmission of a block of data over parallel queues with erasure codes was~\cite{joshi2015queues}. Another analysis focused on the combination of redundant and uncoded requests~\cite{gardner2015reducing}, while a subsequent work analyzed the expected latency of different scheduling policies~\cite{joshi2017efficient}. However, these works still consider simple static queuing policies and only study the one-step latency, neglecting the potential impact of the redundancy on future blocks scheduled on the same queues. Conversely, we examine the long-term effects of the coding and scheduling policy and determine the optimal policy. To the best of our knowledge, these are novel contributions to the theoretical queuing literature.

The scheduling problem over parallel queues is not just theoretical, but a very real issue for multipath transport protocols such as \gls{mptcp}, which can be heavily impacted by the head-of-line blocking problem~\cite{ferlin2014multi}. The most basic \gls{mptcp} scheduler, currently used in the Linux implementation of the protocol, adopts the \gls{lowrtt} policy: packets are sent in the order in which they are written by the application, on the path with the lowest measured \gls{rtt} among those with enough available space in their congestion window. Round robin~\cite{choi2017optimal} and loss-based~\cite{dong2017lamps} scheduling schemes have also been proposed, but fail when there is a strong imbalance between the subflows.  
These heuristics are often inefficient~\cite{hwang2015packet} and can lead to significant performance losses~\cite{ni2014fine}. In fact, it might be convenient to send data on slower paths in advance, exploiting the difference in the path's \glspl{rtt} to have packets arriving in the correct order to the receiver. This more complex scheduling requires to model each path's \gls{rtt} and capacity in order to properly interleave packets among the parallel paths. Schedulers such as the \gls{stms}~\cite{hang2018stms} and \gls{daps}~\cite{kuhn2014daps} are designed to do so, and have better performance than simpler heuristics in most situations. The \gls{blest}~\cite{ferlin2016blest} scheduler adds the awareness of the possibility of head-of-line blocking to this mechanism, explicitly trying to prevent it.

However, standard \gls{mptcp} is not a viable option for real-time traffic, particularly over wireless links ~\cite{chen2013measurement}, makes reliable low-latency service extremely hard to provide. 
Packet-level coding can help to avoid lengthy retransmissions: if a packet is lost on one path, it can be recovered from redundancy packets received on other paths. Several coding-based \gls{mptcp} schedulers have been proposed~\cite{garcia2017low,cui2015fmtcp}, using different coding schemes. \gls{dems}~\cite{guo2017accelerating} is a first attempt to exploit parallel paths to deliver messages from block-based applications: in a two path setting, \gls{dems} transmits data on one path starting from the beginning of the block, and on the other path starting from the end. The scheduler foresees an adaptive redundancy mechanism to improve delivery times in variable network conditions. 
The same concept was the basis of the \gls{hop}~\cite{chiariotti2021hop}, which gives explicit \gls{qos} guarantees by using a greedy policy. These two protocols aim at guaranteeing reliable low-latency communications on a block-by-block basis. For a more thorough survey of the state of the art on scheduling in multipath transport protocols, we refer the reader to~\cite{polese2019survey}.

The theoretical studies on fork-join queuing and the practical work on multipath scheduling have a fundamental limit: while the parallel queue scheduling problem requires foresighted strategies to avoid self-queuing delay and balance the tradeoff between immediate and future reliability, the existing literature focuses on one-step heuristics and policies, using \emph{ad hoc} mechanisms to ensure stability. To the best of our knowledge, our work is the first to provide a rigorous model of the long-term tradeoff, providing the optimal policy even with delayed feedback and packet erasures. Moreover, our model combines most of the features studied in the literature, providing a complete approach to redundant multipath communication.

\section{System Model} \label{sec:sysmodel}

We consider a sender that periodically generates blocks of $K$ packets of $g$ bits every $\tau_g$ seconds. Each block has to be delivered within a deadline $\tau_d$ from its generation time. Vectors are denoted in bold, \glspl{pdf} and \glspl{pmf} are denoted with lower-case letters, whose upper-case versions indicate the respective \glspl{cdf}. The assumption that the traffic source is periodic and has packets with constant size is required for tractability, but is also justified by a number of use cases. Recent technical documents from industrial associations~\cite{TS22104,TS22186} and research papers~\cite{szymanski2016supporting,kuo2019assessment} describe the use of periodic messages of constant size in several industrial scenarios: a recent 3GPP specification~\cite{TS22104} defines the \emph{deterministic periodic communication} class for closed-loop control systems, specifically targeting reliable, low-latency transmission and distinguishing between wireless channel latency and end-to-end latency. While most of these applications are low-throughput, some cases (such as video transmission or high-dimensional sensing data) can run into the tradeoff studied in this work. Another use case in which \gls{cbr} flows play a crucial role is represented by \gls{v2x} communications and cooperative driving~\cite{qi2020traffic}, as specified by the 3GPP~\cite{TS22186}. Finally, while \gls{vr} applications are mostly not \gls{cbr}, some commercial applications use it to provide a smoother service~\cite{lecci2021open}. 

The sender has $M$ available connections to the receiver with independent bottleneck links, using different technologies. The objective of the controller is to encode the $K$ packets of the generic $i$-th block into $N_i$ encoded packets, then schedule them for transmission over the links. The proposed model and schemes are agnostic to the specific implementation of the packet-level code, as long as it allows a block to be decoded as soon as any set of $K$ packets is received. Efficient implementations of systematic packet-level codes with this property, such as shortened-and-punctured Reed-Solomon codes~\cite{rizzo1997effective}, are available in the literature.

The schedule is decided once for the whole block, as the transmitter needs to encode the packets and send them.  The notion of blocks of data that need to be delivered as a whole is natural to most throughput-intensive applications, so including it in the optimization is more effective than packet-by-packet scheduling, which considers individual packets as the scheduling units. In the following, we will use block generation times as a natural timestep for the optimization, as they correspond to scheduling decisions; however, the communication is performed over continuous time, so packets and feedback may be delivered at any point in time. 

The connections are modeled as parallel single-server \glspl{qs}, denoted as $Q_1,\ldots,Q_M$: this model is extremely common in the networking literature, and has historically been used to approximate end-to-end connections~\cite{bonald1999comparison,vinnicombe2002stability}, and more recently, in the optimization of Cloud systems~\cite{gardner2015reducing,joshi2017efficient}. In general, modeling an end-to-end connection as the combination of a stochastic queuing system and a deterministic propagation delay is common when there is a significant bottleneck in the wireless access link~\cite{gholipoor2020cloud,mathew2019packet}, which is the main application scenario we envision for multipath wireless communications. 
We can also consider the state of the queues and channels as \glspl{mc} using the block generation times as a timestep, computing the transition probabilities that correspond to the possible sequences of packet deliveries between subsequent blocks.
The transmitter then divides the $N_i$ encoded packets belonging to block $i$ over the $M$ \glspl{qs}, generating the scheduling vector $\mathbf{s}(i)=(s_1(i),\ldots,s_M(i))$, where $s_m(i)$ denotes the number of packets scheduled for transmission on $Q_m$, with $\sum_{m=1}^M s_m(i)=N_i$. Furthermore, the vector $\mathbf{q}(i)=(q_1(i),\ldots,q_M(i))$ indicates the number of packets in the $M$ queues just before the generation time of the $i$-th block of packets. The service times in the \glspl{qs} are then independent, while the arrival process depends on the scheduler, which takes into account the state of all \glspl{qs}.

\begin{figure}[t]
  \centering
         \input{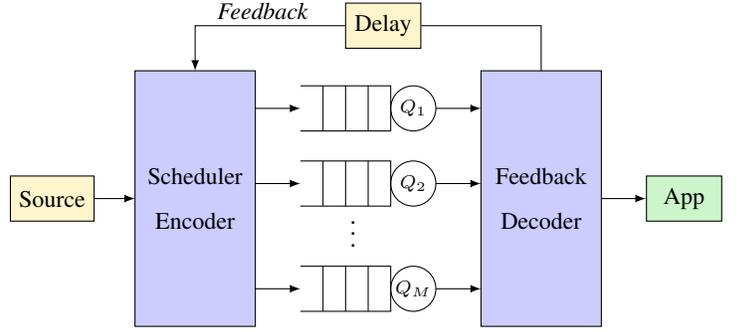}
        \caption{Basic schematic of a fork-join \gls{qs}.}
        \label{fig:sysmodel}
\end{figure}

The block is considered to be delivered whenever at least $K$ of the $N_i$ transmitted packets reach the destination within the deadline, across any combination of \glspl{qs}. We assume that packets sent on each channel $m$ can be randomly and independently dropped with probability $\varepsilon_m$ after service. Such events are named \emph{erasures}, and are assumed to occur the $m$-th \gls{qs} according to an independent Bernoulli process with parameter $\varepsilon_m$. We conservatively assume that erased packets occupy the servers as any other packet, but are discarded by the receiver. Feedback about each delivered or erased packet is assumed to be received by the sender with a constant delay $\tau_f$. If $\tau_f>0$, the sender does not know the state of the queues at time $t$, but only their past state at time $t-\tau_f$. As we will discuss in detail in Sec.~\ref{ssec:delay}, we can still find the optimal schedule given the available information on the system state, but the delay in the knowledge of the state makes the policy slightly more conservative. The case $\tau_f=0$ is valid for local networks in which the sender is co-located with the wireless access point or connected directly to it, so that the delay of the return channel (which is assumed to be lightly loaded) can be assumed to be negligible with respect to that of the forward channel.
A schematic of the system is depicted in Fig.~\ref{fig:sysmodel}, and the main notation used in the paper is listed in Table~\ref{tab:notation}. 

Note that the actual size of each batch depends on the controller's decisions; even though we assume that the blocks are generated according to a deterministic process, each \gls{qs} is a $G/G/1/\ell$ system according to Kendall's notation, where $\ell$ is the (finite) capacity of the queue.
The service time depends, in general, on the queuing state and on the underlying channel state. However, note that the problem is solvable only if the average service rate is higher than the uncoded arrival rate. As multipath communications are usually over different technologies or orthogonal frequency bands, we assume there are no direct interactions between the paths, and connections can be modeled as independent \glspl{qs}. We can consider a general model in which connections are time-varying,
and their bandwidth depends on a parameter $c_m$, which changes according to a \gls{mc} with state space $\mathcal{C}_m= \{ 1,\ldots, C_m\}$ and transition matrix $\bm{\theta}_{m}$ (i.e., there is a finite set of possible values for $c_m$). A classic example is the binary Gilbert-Elliott model~\cite{gilbert1960capacity}. We denote the combination of $c_m(i)$ and $q_m(i)$ as seen by the $i$-th block as $\psi_m(i)=(c_m(i),q_m(i))$.

\begin{table*}[t]
\tiny
	\begin{center}
		\begin{tabular}{cl|cl|cl}
			\toprule
Symbol & Meaning & Symbol & Meaning  & Symbol & Meaning  \\	\midrule
$K$ & Size of a data block (in packets) & $M$ & Number of \glspl{qs} & $c_m(i)$ & \gls{qs} state\\
$\tau_d$ & Delivery deadline & $\tau_g$ & Inter-block generation time& $\bm{\Theta}_m$ & \gls{qs} state transition matrix\\
$q_m(i)$ & Queue state & $\psi_m(i)$ & Overall state for \gls{qs} $m$ seen by block $i$ & $\mathcal{Y}_{c_m(i)}(n)$ & $n$-th packet service time in state $c_m(i)$\\
$\mu_m$ & \gls{qs} average service rate & $\ell$ & Queue length (in packets) & $\varepsilon_m$ & Packet erasure probability\\
$T$ & Transition matrix & $\mathcal{A}$ & Action set & $\mathbf{s}$ & Scheduling vector \\
$\tau_f$ & Feedback delay & $\mathcal{D}_m$ & Delivered packets before the next block & $y_{\psi_m(i)}(n)$ & Overall delay of the $i$-th packet\\
  $\rho$ & Block delivery probability & $\delta_m$ & \gls{pmf} of the total number of delivered packets & $\omega_m$ & \gls{pmf} of the number of delivered packets\\
$\mathcal{N}(\mathbf{s})$ & Set of delivery vectors leading to decoding & $\Pi$ & Scheduling policy & $\bm{\phi}$ & Steady-state distribution  \\
$r_{\bm{\psi}}$ & Reward function & $\lambda$ & Discount factor & $R_{\bm{\psi}}$ & Long-term reward\\
$\Gamma$ & \gls{cdf} of the block delivery delay & $\tilde{\bm{\psi}}$ & Delayed state &&\\
\bottomrule
		\end{tabular}
	\end{center}
\caption{Main notation used in the paper.}
	\label{tab:notation}
\end{table*}

For the generic $m$-th \gls{qs}, the service time of the $n$-th packet transmitted on that \gls{qs} after the generation of block $i$ is then a random variable $\mathcal{Y}_{c_m(i)}(n)$, which is conditioned on $c_m(i)$. In this section, we do not make any assumptions about the distribution of the service time, so as to maintain full generality. The $M$ \glspl{qs} are mutually independent, so that $\psi_m(i)$ and $\mathcal{Y}_{c_m(i)}(n)$ are independent between the \glspl{qs}. We also define the  state vector $\bm{\psi}(i) =(\psi_1(i), ...,\psi_m(i))$, containing the states of the $M$ \glspl{qs} for block $i$.

Finally, we define $\mu_m(i) = \frac{1}{\mathbb{E}\left[\mathcal{Y}_{c_m(i)}\right]}$ as the average service rate of $Q_m$ for block $i$: by the fundamental renewal theorem, the service rate can be computed as the number of packets in a renewal interval, i.e., 1 (as the service of 1 packet corresponds to a renewal) divided by the average duration $\mathbb{E}\left[\mathcal{Y}_{c_m(i)}\right]$. We can extend this to get $\mu(i) = \sum_{m = 0}^{M} \mu_m(i)$, the aggregate service rate of all the \glspl{qs}. We make the conservative approximation (which is exact in the $G/M/1/\ell$ case) that, upon a new block arrival to a \gls{qs}, the residual service time of the packet under service is distributed as a general service time (i.e., we neglect the time already spent in service). As we show in the Appendix, the actual probability of the packet being delivered before a given time is always lower than the approximated value for light-tailed service time distributions, i.e., distributions with a monotonically non-decreasing hazard rate~\cite{barlow1963properties}. The resulting policy will be suboptimal in the actual system, as it will consider a slightly different state, but this only affects the first packet in the queue, and as such, has a limited effect on the system for larger blocks. 

\section{MDP Formulation and Solution}\label{sec:mdp}

We can now model the scheduling process as a finite \gls{mdp}, whose decision instants correspond to the arrival of new blocks. \glspl{mdp} are defined by a state space, an action space, a matrix of transition probabilities, and a reward function. The state of the \gls{mdp} contains all the information available to the controller when it makes a decision, while the actions are naturally the possible schedules that can be applied, and the reward is the success probability of current and future blocks. In our case, the state of the system for block $i$ is $\bm{\psi}(i)$. Consequently, the state space is simply $\Psi=\left(\prod_{m=1}^M\mathcal{C}_m\right)\times\{0,\ldots,\ell\}^M$. In the following, we first model the case in which feedback is instantaneous, i.e., $\tau_f=0$, then extend the derivation to the general case.

The action space for the controller is also simple. Since the number of packets on each \gls{qs} cannot exceed the queue capacity $\ell$, the action space is simply $\mathcal{A}=\{0,\ldots,\ell\}^M$, and each action is a possible vector $\mathbf{s}(i) \in \mathcal{A}$. It should be noted that any actions that entail transmitting fewer than $K$ packets in total are always outperformed by dropping the block entirely (i.e., $\bm{s}(i) = \bm{0}$, also referred to as a \emph{block drop}), and can then be removed from the action set.

The transition probabilities $T(\bm{\psi}, \bm{\psi}',\mathbf{s})$ can be computed from the statistics of the service time. The delay $y(n|\psi_m(i))$ for the delivery of the $n$-th scheduled packet on \gls{qs} $m$ is given by:
\begin{equation}
 y(n|\psi_m(i))=\sum_{j=1}^{q_m(i)+n}\mathcal{Y}_{c_m}(j).
\end{equation}
The \gls{pmf} of $y(n|\psi_m(i))$, which accounts for the time to serve the residual $q_m(i)$ packets of the previous blocks (if any), plus the time to serve the $n$ packets scheduled on queue $m$ for the current block, can be  computed from the known statistics of the service time.
In turn, the \gls{pmf} $\delta_m(x|\tau , \psi_m,s_m)$ of the number of packets $\mathcal{D}_m$ delivered by time $\tau$ can be computed as:
\begin{equation}
\begin{small}
\delta_m(x| \tau , \psi_m,s_m) = \begin{cases}
 P \left[y(1|\psi_m(i))>\tau\right], &   x=0;\\[5pt]
 \begin{split}
 &P [y(x|\psi_m(i))\leq\tau, \\ &y(x+1|\psi_m(i))>\tau ],
 \end{split} & 0 <x < q_m+s_m; \\[10pt]
 P \left[y(x|\psi_m(i))\leq\tau\right], &   x = q_m+s_m,
 \end{cases}\label{eq:deliverycdf}
 \end{small}
\end{equation}
and 0 in all other cases. We can easily derive the \gls{cdf} of $\mathcal{D}_m$, $\Delta_m(x| \tau , \psi_m,s_m)$, as:
\begin{equation}
\Delta_m(x| \tau , \psi_m,s_m) = \sum_{k=0}^x\delta_m(k| \tau , \psi_m,s_m).
\end{equation} 

\subsection{Reward Calculation}\label{ssec:reward}

The reward function will depend on the probability that the block of data is decoded within its deadline $\tau_d$. In this case, we also have to consider packet erasures. In order to deliver new packets, a \gls{qs} must first flush the queue, i.e., deliver the $q_m(i)$ packets already in flight (irrespective of whether they are erased, as they do not count for the current block). Let $\omega_m(x|\tau,\varepsilon_m , \psi_m,s_m)$ be the \gls{pmf} of the number of packets $\omega_m$ of the current block useful for the block reconstruction that were received from \gls{qs} $m$ over the time interval $\tau$. We hence have
\begin{multline}
 \omega_m(x|\tau,\varepsilon_m ,\psi_m,s_m)=\sum_{r=x}^{s_m}\delta_m(r+q_m| \tau , \psi_m,s_m) \\\binom{r}{r-x}\varepsilon_m^{r-x}(1-\varepsilon_m)^{x},\label{eq:correct_del_cdf}
\end{multline}
where the rightmost term accounts for the probability that $x$ packets out of $r$ are delivered, while $r-x$ are erased (we remind the reader that packet erasures on link $m$ occur independently with probability $\varepsilon_m$).
The \gls{cdf} $\Omega_m(x|\tau,\varepsilon_m , \psi_m,s_m)$ of the number of packets delivered in time is obtained by simply summing the \gls{pmf} in~\eqref{eq:correct_del_cdf}.
Extending the calculation from a single \gls{qs} to the overall system, the block delivery \gls{pmf} is given by the convolution of the  \glspl{qs}' \glspl{pmf}:
\begin{equation}
 \rho(\tau, \bm{\varepsilon} \vert \bm{\psi},\mathbf{s})=\sum_{\mathbf{n}\in\mathcal{N}(\mathbf{s})}\prod_{m=1}^M \omega_m(n_m|\tau, \varepsilon_m , \psi_m,s_m).\label{eq:rho}
\end{equation}
where $\mathcal{N}(\mathbf{s})$ is the set of all possible vectors $\mathbf{n} = (n_1, ...,n_M)$ of correctly received packets that result in the block being successfully decoded, i.e.:
\begin{equation}
\begin{split}
 \mathcal{N}(\bm{s})=\bigg\{\mathbf{n}\in\mathcal{A}:n_m\leq s_m, \forall m\in\{1,\ldots,M\}, \\ \sum_{m=1}^M n_m\geq K\bigg\}.
 \end{split}
\end{equation}
In this work, we choose the reward function $r_{\bm{\psi}}(\mathbf{s}) = \rho(\tau, \bm{\varepsilon} \vert \bm{\psi},\mathbf{s})$.
We now define a \emph{policy} $\Pi:\Psi\rightarrow\mathcal{A}$ mapping states to actions. Let $\mathbf{r}(\Pi)$ be the vector with elements $r_{\bm{\psi}}(\Pi(\bm{\psi}))$ associated to the states $\bm{\psi}\in\Psi$, and let $\mathbf{T}_{\Pi}$ be the transition probability matrix associated with the policy, whose elements are simply given by $ T_{\Pi}(\bm{\psi},\bm{\psi}', \tau_g)=T(\bm{\psi},\bm{\psi}'|\Pi(\bm{\psi}), \tau_g)$.
We can define the long-term discounted reward $R_{\bm{\psi}}(\Pi)$ from state $\psi$ as
\begin{equation}
 R_{\bm{\psi}}(\Pi)=\sum_{j=0}^\infty\lambda^j \sum_{\bm{\psi}'\in\Psi}P[\bm{\psi}'(j)=\bm{\psi}'|\bm{\psi}(0)=\bm{\psi};\Pi]r_{\psi'}(\Pi),
\end{equation}
where $\lambda\in[0,1)$ is the discount factor. The probability of being in state $\bm{\psi'}$ after $j$ steps is an element of the matrix $\mathbf{T}_{\Pi}$ elevated to the $j$-th power. The vector $\mathbf{R}(\Pi)$, whose elements are associated to the long-term reward for each state for the given policy, is then such that $\mathbf{R}(\Pi) = (\mathbf{I} - \lambda \mathbf{T}_{\Pi})^{-1} \mathbf{r}(\Pi)$.
This allows us to compute the total reward starting from the known $\mathbf{r}(\Pi)$ and $\mathbf{T}_{\Pi}$.
Finally, from the transition matrix we can also compute the steady state probability distribution $\bm{\varphi}$ of the system, and therefore the steady-state total reward $R(\Pi) = \bm{\varphi}^T \mathbf{R}(\Pi)$, where $(\cdot)^T$ is the transpose operator.
Moreover, recalling~\eqref{eq:rho} and defining the vector $\bm{\rho}(\tau, \Pi, \epsilon)$ as the vector of values $\rho(\tau,\bm{\varepsilon}|\bm{\psi}, \Pi(\bm{\psi}))$ over the whole state space, the \gls{cdf} of the block delivery time can be expressed as $\Gamma(\tau) = \bm{\varphi}^T \bm{\rho}(\tau, \Pi, \epsilon)$.
\subsection{Delayed Feedback}\label{ssec:delay}

We now consider the delayed feedback scenario, in which $\tau_f>0$. In this case, the state of the \gls{mdp} is not the real queue size, but the size of the queue as perceived by the sender, i.e., with a delay of $\tau_f$, and the state of the \gls{mc} driving the service times is delayed by one step. We denote the delayed state as $\tilde{\bm{\psi}}$.
The \gls{cdf} of the number of delivered packets on a \gls{qs}, given in ~\eqref{eq:deliverycdf}, needs to be updated to reflect the delay in the feedback:
\begin{multline}
\tilde{D}_m(x|\tau_d,\tau_f , \tilde{\bm{\psi}},s_m)=\sum_{c_m'\in\mathcal{C}_m}\bigg[\theta_m(\tilde{c}_m,c_m')\cdot \\ \sum_{r=0}^{\tilde{q}_m}  \delta_m\left(r|\tau_f , \tilde{\psi}_m,0\right) \Delta_m\left(x|\tau_d , \left(c_m',\tilde{q}_m-r\right),s_m\right) \bigg].\label{eq:delaydeliverycdf}
\end{multline}
This equation takes into account that packets that have not been acknowledged might already have been delivered, computing the \emph{a posteriori} probability of the real state $\bm{\psi}$ given the delayed observation $\tilde{\bm{\psi}}$.
The values of $\tilde{\Omega}_m(x|\tau ,\varepsilon_m , \tilde{\psi}_m, s_m)$ and $\tilde{\rho}(\tau, \bm{\varepsilon} \vert \bm{\psi},\mathbf{s})$ can then be calculated with the same method we used in~\eqref{eq:correct_del_cdf} and~\eqref{eq:rho}, after simply replacing ~\eqref{eq:delaydeliverycdf} to ~\eqref{eq:deliverycdf}. 

The transition probabilities for each \gls{qs} also need to be updated as follows:
\begin{equation}
\begin{small}
\begin{aligned}
 \tilde{T}(\tilde{\bm{\psi}},\tilde{\bm{\psi}}'|\mathbf{s},\tau_g,\tau_f)=\prod_{m=1}^M\sum_{r=0}^{\tilde{q}_m}\bigg[\theta(c_m,c_m')\delta_m\left(r|\tau_f , (c_m,0),\tilde{q}_m\right)\\
 \times \delta_m \left(\tilde{q}_m-r+s_m-\tilde{q}_m'| \tau_g - \tau_f , (c_m',0),\tilde{q}_m-r+s_m \right) \bigg].
 \end{aligned}
\end{small}
\end{equation}
The equations from Sec.~\ref{ssec:reward} can be used to find the optimal policy with delayed feedback, replacing $\mathbf{T}_m$, $D_m$, $\Omega_m$, and $\rho$ with their \emph{a posteriori} delayed versions $\tilde{\mathbf{T}}_m$, $\tilde{D}_m$. $\tilde{\Omega}_m$, and $\tilde{\rho}$.

\subsection{Computation of the Optimal Policy}\label{ssec:pi}

We can now solve the problem using the classic policy iteration algorithm~\cite[Ch. 4]{sutton2018reinforcement}, which consists of two steps, policy evaluation and policy improvement, which are repeated until convergence. The algorithm is initialized with a policy function $\Pi^0$ and a value function $v_\Pi^0$, which are both set to all zeros. The iterative steps are then the following:
\begin{enumerate}
    \item The policy is evaluated using \begin{equation}
    \begin{split}
        v_\Pi^{n+1}(\psi)=\sum_{\psi'\in\Psi}p(\psi'\vert \psi,\Pi^n(\psi)) \\ \left(r(\psi,\Pi^n(\psi),\psi')+\lambda v^n_\Pi(\psi')\right),
        \end{split}
    \end{equation}
     for all $\psi$, where $\psi$ is the current state, $\psi'$ is the new state, $\mathbf{s}$ is the chosen schedule, and $r$ is the instantaneous reward. The value function is an estimate of the long-term value that can be achieved in a given state using policy $\Pi^n$.
    \item The policy is improved by choosing the action that maximizes the long-term value:
    \begin{equation}
    \begin{split}
        \Pi^{n+1}(\psi)=\argmax_{\mathbf{s}\in\mathcal{A}} \sum_{\psi'\in\Psi}p(\psi'\vert \psi,\mathbf{s}) \\ \left(r(\psi,\mathbf{s},\psi')+\lambda v^{n+1}_\Pi(\psi')\right).
        \end{split}
    \end{equation}
\end{enumerate}
Policy iteration is guaranteed to converge to the optimal policy in finite-state \glspl{mdp} with finite reward, as it improves the policy at each step, gradually converging to the optimum~\cite{howard1960dynamic}. In the general case, the complexity of policy iteration is exponential in the number of states, making it particularly impractical for realistic problems. However, if the correct pivoting rule is adopted and the discount factor is constant in time,  policy iteration reaches convergence after $N_{\text{it}}\leq\frac{|\Psi|^2(|\mathcal{A}|-1)}{1-\lambda}\log\left(\frac{|\Psi|^2}{1-\lambda}\right)$ cycles, and is strongly polynomial in the size $|\Psi|$ of the state space and the size $|\mathcal{A}|$ of the action space~\cite{ye2011simplex}.
Each iteration uses at most $O(|\mathcal{A}||\Psi|^2)$ operations, making the resulting bound polynomial in the \gls{mdp} size.
Naturally, due to the curse of dimensionality, even a two-path system is extremely complex in practice, as there are thousands of possible states and tens of actions. 
It is worth remarking that the model can be extended to account for more general assumptions, such as random block size distributions, different service time statistics for the $M$ queues, other coding schemes, and diverse queue lengths. However, such generalizations take a toll in terms of complexity of notation and analysis. We hence preferred simplicity over generality, in an effort to make it easier for the reader to follow the rationale and capture the essence of the proposed study. In any case, the complexity of the policy iteration approach can soon become an obstacle in practical settings, because of the exponential growth of the computational complexity. Reinforcement learning solutions might be a practical alternative to policy iteration if the problem becomes too large, but we leave this analysis to future work.

\section{Analytical policy derivation}\label{sec:analytical}

For very limited cases, it is possible to derive an analytical solution to the problem by computing the expected reward for each policy without recurring to iterative strategies. In particular, we consider a system with unit block size (i.e., $K=1$) and $M=2$ \glspl{qs} with queue capacity $\ell=1$: if a packet is already in service, no further packets can be queued in the same \gls{qs}. We consider exponential service times with rates $\mu_1$ and $\mu_2$. Without loss of generality, we also assume $\mu_1 \geq \mu_2$. The \glspl{qs} have instantaneous feedback and no erasures. In this case, there are only 4 possible states, namely, $\psi \in \{0,1\}^2$. For the state $(1,1)$ (i.e., when both queues are occupied), the only meaningful action is to transmit no packets, as anything that is transmitted will be dropped by the \gls{qs}.
Similarly, transmitting packets on the first link in state $(1,0)$ is meaningless, so the only meaningful actions in that state are $\mathcal{A}((1,0)) \in \{(0,0),(0,1)\}$, and the same goes for state $(0,1)$, with swapped indices. We can also immediately discard action $(0,0)$ in state $(0,0)$, which corresponds to the system never transmitting anything, so the meaningful actions in state $(0,0)$ are $\{(0,1), (1,0), (1,1)\}$.

\begin{table}[t]
\tiny
	\begin{center}
		\begin{tabular}{ c | cccccccc}
			\toprule
\multirow{2}{*}{State} & \multicolumn{8}{c}{Policy}\\
& A & B& C & D & E & F & G & H\\
\midrule
(0,0) & (1,1) & (1,1)& (1,1)& (1,1)& (1,0)& (1,0)& (0,1)& (0,1)\\
(1,0) & (0,1)& (0,1)& (0,0)& (0,0)& (0,1)& (0,1)& (0,1)& (0,0)\\
(0,1) & (1,0)& (0,0)& (1,0)& (0,0)& (1,0)& (0,0)& (1,0)& (1,0)\\
(0,0) & (0,0)& (0,0)& (0,0)& (0,0)& (0,0)& (0,0)& (0,0)& (0,0)\\
 \bottomrule
		\end{tabular}
	\end{center}
\caption{Possible strategies.}
	\label{tab:strats_analytical}
\end{table}

With this knowledge, and excluding policies which differ only for actions in unreachable states and policies that only use one channel, we can identify the $8$ non-trivial policies listed in Table~\ref{tab:strats_analytical}. We also note that the results for symmetric policies can be easily computed by swapping the indices and parameters. We have an \gls{mdp} with the transition probability matrix $\mathbf{T}$ for a generic policy $\Pi$ and expected instantaneous reward $\rho(q_1,q_2)$ expressed in equations ~\eqref{eq:tmatrix_analytical} and ~\eqref{eq:reward_analytical} (see pg. \pageref{eq:analytical}) where $p_m=e^{-\mu_m\tau_g}$ and $r_m=e^{-\mu_m\tau_d}$.
The expected reward is $\mathbb{E}[\rho|\Pi]=\bm{\phi}_{\Pi}\bm{\rho}_{\Pi}$, where the steady-state distribution $\bm{\phi}_{\Pi}$ is the left eigenvector of matrix $\mathbf{I}-\mathbf{T}_{\Pi}$ with eigenvalue 1.

\begin{figure*}[t]
\begin{equation}
 \mathbf{T}_{\Pi}=\begin{tiny}\left(
\begin{array}{cccc}
\left(1-p_1\Pi_1(0,0)\right)\left(1-p_2\Pi_2(0,0)\right) & p_1\Pi_1(0,0)\left(1-p_2\Pi_2(0,0)\right) & p_2\Pi_2(0,0)\left(1-p_1\Pi_1(0,0)\right) &  p_1\Pi_1(0,0) p_2\Pi_2(0,0)\\
\left(1-p_1\right)\left(1-p_2\Pi_2(1,0)\right) & p_1\left(1-p_2\Pi_2(1,0)\right) & p_2\Pi_2(1,0)\left(1-p_1\right) & p_1p_2\Pi_2(1,0)\\
\left(1-p_1\Pi_1(0,1)\right)\left(1-p_2\right) & p_2\left(1-p_1\Pi_1(0,1)\right) & p_1\Pi_1(0,1)\left(1-p_2\right) &  p_1\Pi_1(0,1)p_2\\
(1-p_1)(1-p_2) & p_1(1-p_2) & p_2(1-p_1) & p_1p_2\\
\end{array}
\right) \end{tiny}\label{eq:tmatrix_analytical}
\end{equation}
\begin{equation}
\normalsize\bm{\rho}_{\Pi}\!=(1-(1-\Pi_1(0,0)(1-r_1))(1-\Pi_2(0,0)(1-r_2)), \Pi_2(1,0)(1-r_2), \Pi_1(0,1)(1-r_1), 0)^T, \label{eq:reward_analytical}
\end{equation}
\hrule \label{eq:analytical}
\end{figure*}

As an example, we analyze policy A, which always transmits when possible. Its steady-state probability is $\bm{\phi}_A=\left((1-p_1)(1-p_2), p_1(1-p_2), p_2(1-p_1), p_1p_2\right)$. The expected reward is then:
\begin{equation}
\begin{split}
 \mathbb{E}_A[\rho]=\bm{\phi}_A\bm{\rho}_A=&1-p_1p_2-p_2r_1(1-p_1)-p_1r_2(1-p_2)\\&-r_1r_2(1-p_1)(1-p_2).
 \end{split}
\end{equation}

In general, finding the optimal policy requires an enumeration over the policy space, but this is computationally light, as the expected reward can be computed in closed form as we did for policy A.
In the case where $\mu_1=\mu_2=\mu$, and consequently $p_1=p_2=p$ and $r_1=r_2=r$, the solution is simple, and the optimal policies are only $A$ or $E$ (which is identical to $F$, as the two paths have the same rate in this case). The expected reward for policy $E$ is given by:
\begin{equation}
 \mathbb{E}_E[\rho]=(1-p)(1-r)-\frac{p(1-p)(1-r)}{1+p(1-p)}.
\end{equation}
Which policy is optimal then depends on the precise values of the parameters $\tau_g$, $\tau_d$, and $\mu$, and we can define a boundary function that tells us the region of the parameter space in which the optimal policy is $A$, i.e., the values of the parameters for which $\mathbb{E}_A[\rho]\geq\mathbb{E}_E[\rho]$:
\begin{equation}
\tau_g\geq\frac{1}{\mu}\ln\left(\frac{1+\sqrt{4e^{\mu\tau_d}-3}}{2}\right)\Rightarrow\Pi^*=A.
\end{equation}

\begin{figure}[t]
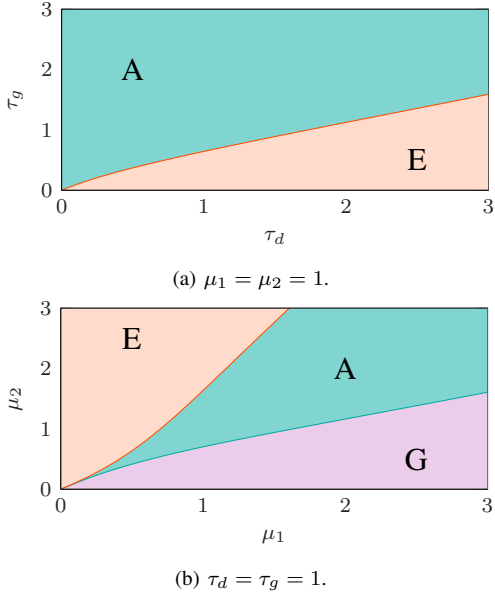

     \centering
     \begin{subfigure}[b]{0.40\textwidth}
         \centering
         \input{tikz_figures/revision/analytical_constant_mu.tex}
         \caption{$\mu_1=\mu_2=1$.}
         \label{fig:const_mu}
     \end{subfigure}
     \hspace{0.4cm}
     \begin{subfigure}[b]{0.40\textwidth}         
         \centering
         \input{tikz_figures/revision/analytical_constant_tau.tex}
         \caption{$\tau_d=\tau_g=1$.}
         \label{fig:const_tau}
     \end{subfigure}
        \caption{Optimal action boundary in the $\ell=1$ example.}
        \label{fig:analytical}
\end{figure}

Fig.~\ref{fig:analytical} shows the optimal policy as a function of the parameter values in the cases with $\mu_1=\mu_2=1$ and $\tau_g=\tau_d=1$. In the first case, policies E and G are equivalent, as the two paths are identical. Asymmetrical policies such as E and G, which privilege one of the two paths and use the other as backup, are more convenient in cases in which the two paths have very different rates, while the symmetrical policy is optimal if the two paths are similar. 

The set of possible deterministic policies contains $|\mathcal{A}|^{|\mathcal{S}|}$ elements, as a policy is a function $\Pi:\mathcal{S}\rightarrow\mathcal{A}$ mapping states to actions. Computing the expected reward for each policy explicitly, as we did for the simple example, has exponential complexity in terms of the state space size $|\mathcal{S}|$, while policy iteration converges to the optimum in polynomial time~\cite{ye2011simplex}. Therefore, we will only consider policy iteration for the full-sized problem, as both methods return the optimal policy, but the iterative solution is practically computable in a short time.

\section{Results}\label{sec:results}
In this section, we investigate the performance and behavior of the optimal and heuristic strategies in some specific scenarios. We choose to define $r_{\psi}(\mathbf{s})$ as the delivery probability itself and the discount factor as $\lambda = 0.99$. With this choice the long term reward is between $0$ and $100$ and is a linear function of the delivery probabilities. As the discount factor is very close to 1, the reward is close to the long-term probability of success, expressed as a percentage, with only a slight preference for immediate rewards. We derived the optimal policy in each case using policy iteration, which converged in fewer than 10 iterations in all cases.

All the results below are derived analytically by computing the steady-state probabilities of the \gls{mdp} and applying the strategies in each scenario. We can now analyze the optimal policy for the fork-join system in different conditions.

\subsection{Heuristic Policies} \label{sec:heurPolicy}
We can now look at some practical policies, which are implemented in real \gls{mptcp} schedulers.

The \emph{\gls{ccr}} policy sets a constant amount of redundancy $\beta\in[1,2]$, such that $N=\beta K$ ($\beta$ is the inverse of the coding rate). The $N$ packets are then split among the \glspl{qs} proportionally to their rate at the current state. The policy can thus be defined as:
\begin{equation}
s_m(i) = \left \lfloor \frac{\mu_m(i) N }{\mu(i)} + \frac{1}{2} \right \rfloor.
\end{equation}
The \gls{daps}~\cite{kuhn2014daps}, \gls{dems}~\cite{guo2017accelerating}, and \gls{blest}~\cite{ferlin2016blest} schedulers all employ different variants of the \gls{ccr} policy, scheduling packets in different ways but setting a fixed redundancy. The \gls{ccr} policy used in the simulation is an upper bound to their performance, as it finds the optimal scheduling for the given redundancy level.
A particular instance of this policy is that with $\beta=1$, where the original packets are split between the \glspl{qs} without any coding. This policy, which we call \emph{\gls{ps}}, is almost universally used in legacy schedulers such as \gls{stms}~\cite{hang2018stms} or \gls{lowrtt}.

The \emph{greedy} policy aims to achieve a delivery probability of the blocks above a specified threshold. In particular, for each block $i$ the policy $s(i)$ is such that: \emph{(i)}, the schedule is stable, i.e., $s_m(i) \leq \gamma \frac{\tau_i}{\mu_m(i)}$, with $\gamma < 1$, \emph{(ii)}, the schedule should be such that the delivery probability for that block is not less than $P_{\text{thr}}$ if allowed by the first constraint, otherwise it should maximize the delivery probability, and \emph{(iii)}, the schedule should minimize $N$ while satisfying the first two constraints.

This policy can be computed iteratively adding packets to the \gls{qs} that has the highest delivery probability at each iteration, while checking the constraints and total delivery probability, and is used in the \gls{leap}~\cite{chiariotti2019analysis} and \gls{hop}~\cite{chiariotti2021hop} schedulers. It is much more computationally efficient than the optimal policy, as it does not require the full solution of the Bellman equation, but it can also lead to reliability collapse if the \glspl{qs} cannot support the requirements: in that case, indeed, it will progressively increase the redundancy (making the system unstable) until the maximum queue size is reached. On the other hand, the optimal policy can sacrifice some reliability or even drop a block to ensure that future blocks have the best chance to be delivered.

\subsection{Performance Evaluation}
 
 \begin{table}[t]
\tiny
	\begin{center}
		\begin{tabular}{rcccccccc}
			\toprule
			Scenario & $\ell$ & $K$ & $\tau_g$ & $\tau_d$ & $\epsilon$ & $\beta$ & $\gamma$ & $P_{\text{thr}}$ \\ \midrule
			Low load & 60 & 20 & 20 & 12 & 0 & 1.5 & 0.8 & 0.9\\
			Average load & 60 & 20 & 15 & 15 & 0 & 1.3 & 0.8 & 0.9\\
			High load & 60 & 20 & 12 & 20 & 0 & 1.1 & 0.8 & 0.9\\
			\bottomrule
		\end{tabular}
	\end{center}
\caption{Parameters for the analyzed scenarios.}
	\label{tab:exp1}
\end{table}

First, we consider how asymmetries in the capacity of the \glspl{qs} affects the delivery probability. We consider at first a scenario with two \glspl{qs}, each providing \gls{iid} service times with exponential distribution $\mathcal{Y}_m \sim \text{Exp}(\mu_m)$. We assume no delay in the feedback and no erasures. We set $\mu_1 = 1-\alpha$ and $\mu_2 = 1+\alpha$, so that the aggregate rate remains constant when varying the asymmetry parameter $\alpha$ in $[0,1]$\footnote{Note that the two \glspl{qs} are perfectly balanced for $\alpha = 0$ while the larger $\alpha$ the bigger the capacity of $Q_2$ over $Q_1$.}. In the following, all latencies and time intervals are normalized to the average packet service time, which is hence set to 1 time unit.  We define three scenarios, whose main parameters are listed in Table~\ref{tab:exp1}:
\begin{itemize}
 \item The \emph{high load} scenario assumes a offered load (before adding redundancy) of about 83\% of the average capacity: the block generation period is short and the system is always at risk of building up queues. In this case, the latency requirement is relatively relaxed.
  \item The \emph{average load} scenario has an offered load of about 67\%: blocks have a longer generation period, but the deadline is tighter.
 \item The \emph{low load} scenario has an offered load of 50\%, leaving the queues mostly empty if no \gls{fec} is added, but the deadline in this case is extremely tight.
\end{itemize}
We can expect higher coding rates to be highly beneficial in the low load scenario, as in that case the main issue is not queuing delay but the natural variability of the \glspl{qs}. Conversely, coding can be detrimental in the high load scenario, in which queuing is the most pressing issue, and adding redundant traffic to the already significant load on the system can move it closer to instability.

\begin{figure*}[t]
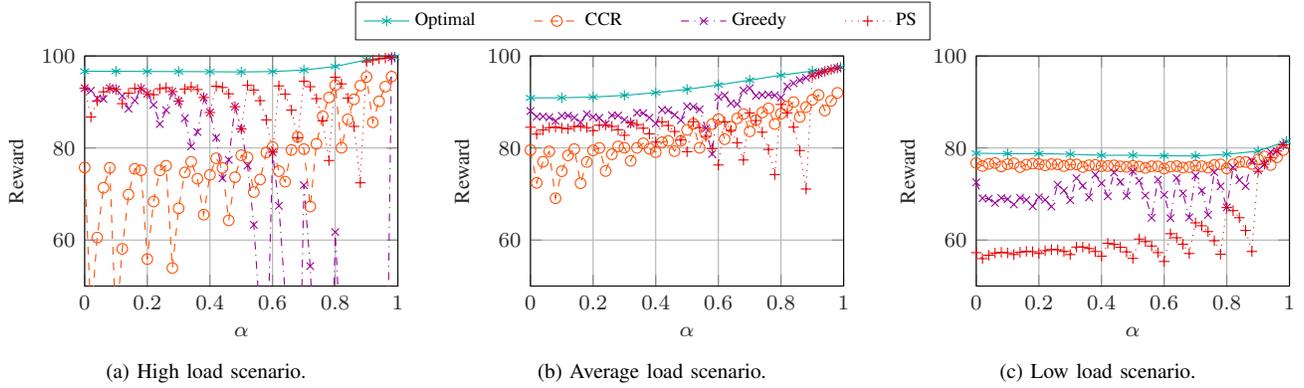

     \centering
     \begin{subfigure}[b]{0.9\textwidth}
         \centering
         \input{tikz_figures/legend_algos}
         \label{fig:legend}
     \end{subfigure}
     \begin{subfigure}[b]{0.32\textwidth}
         \centering
         \input{tikz_figures/scenario2}
         \caption{High load scenario.}
         \label{fig:high_load}
     \end{subfigure}
     \begin{subfigure}[b]{0.32\textwidth}         
         \centering
         \input{tikz_figures/scenario1}
         \caption{Average load scenario.}
         \label{fig:mid_load}
     \end{subfigure}
     \begin{subfigure}[b]{0.32\textwidth}
         \centering
         \input{tikz_figures/scenario3}
         \caption{Low load scenario.}
         \label{fig:low_load}
     \end{subfigure}
        \caption{Reward as a function of the system asymmetry.}
        \label{fig:asymmetry}
\end{figure*}

\subsubsection{Channel Asymmetry Effects} \label{sec:heursitciasymmetry}

The results, showing the achievable reward as a function of $\alpha$, are shown in Fig.~\ref{fig:asymmetry}. In all scenarios we can see that the optimal reward slightly improves when $\alpha \rightarrow 1$, i.e., one channel has a much higher capacity than the other. This is due to the well-known queuing theory result that states that a single \gls{qs} with service rate equal to 2 is better than two parallel ones with service rate 1. We can also observe that the optimal reward varies smoothly with $\alpha$, whereas the other methods presents significant instability due to the finite granularity of packets. 
It is interesting to note that the \gls{ccr} policy performs extremely well, almost at the level of the optimal policy, in the low load scenario, while it is the worst option in the high load scenario: this is because setting a constant level of redundancy can be beneficial if the load is low, but it increases the queuing delay if the load is the main limiting factor, even when adapting the coding rate $\beta$ to the scenario. 
As expected, the \gls{ps} policy shows the opposite pattern, with good performance in the high load scenario. The greedy policy is the closest to the optimum in the average load scenario, as its adaptive nature can balance queue accumulations on the two \glspl{qs}. However, we can see that the high load scenario has a ``snowball effect'' if the \glspl{qs} are asymmetrical: an increase of the queue on the fast \gls{qs} cannot be compensated by the other, which in turn leads the controller to allocate even more redundancy to the fast \gls{qs}, until the system is limited by the stability constraint. Something similar happens in the low load scenario, as the greedy policy is far from its target of 90\% reliability, and will therefore tend to add too much redundancy and cause self-queuing delays.

\begin{figure*}[h]
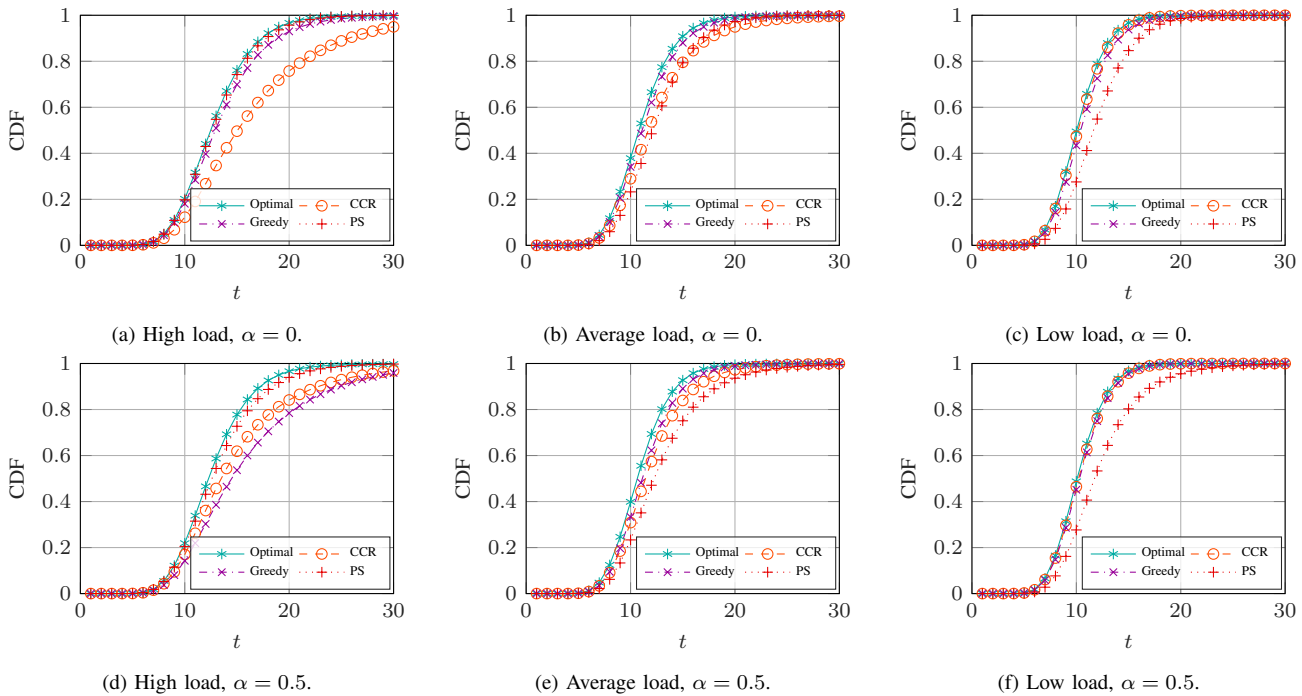

     \centering
        \begin{subfigure}[b]{0.32\textwidth}
         \centering
         \input{tikz_figures/cdf_s2_a0}
         \caption{High load, $\alpha=0$.}
         \label{fig:cdf_high_a0}
     \end{subfigure}
     \begin{subfigure}[b]{0.32\textwidth}
         \centering
         \input{tikz_figures/cdf_s1_a0}
         \caption{Average load, $\alpha=0$.}
         \label{fig:cdf_mid_a0}
     \end{subfigure}
    \begin{subfigure}[b]{0.32\textwidth}
         \centering
         \input{tikz_figures/cdf_s3_a0}
         \caption{Low load, $\alpha=0$.}
         \label{fig:cdf_low_a0}
     \end{subfigure}
    \begin{subfigure}[b]{0.32\textwidth}
         \centering
         \input{tikz_figures/cdf_s2_a05}
         \caption{High load, $\alpha=0.5$.}
         \label{fig:cdf_high_a05}
     \end{subfigure}
    \begin{subfigure}[b]{0.32\textwidth}
         \centering
         \input{tikz_figures/cdf_s1_a05}
         \caption{Average load, $\alpha=0.5$.}
         \label{fig:cdf_mid_a05}
     \end{subfigure}
    \begin{subfigure}[b]{0.32\textwidth}
         \centering
         \input{tikz_figures/cdf_s3_a05}
         \caption{Low load, $\alpha=0.5$.}
         \label{fig:cdf_low_a05}
     \end{subfigure}
        \caption{Delivery time \glspl{cdf}.}
        \label{fig:cdfs123}
\end{figure*}

The \glspl{cdf} of the delivery time in the three scenarios are depicted in Fig.~\ref{fig:cdfs123}. We can see that the \glspl{cdf} for the optimal policy depend mostly on the load of the system. Furthermore, the \glspl{cdf} for the strategies that perform well in each scenario do not have significant differences in shape, which suggests that they are robust to changes in the deadline and reliability threshold. As we discussed above, the load on the system is the most important parameter in determining the efficiency of the heuristic schemes, as well as the optimal performance.

\subsubsection{Markov-Modulated Channels}\label{sec:markovoptimal}

We now examine what happens when \glspl{qs} have variable capacities. In the Markov scenario the service time, as a function of the asymmetry parameter $\xi$, is distributed as $\mathcal{Y}_{c_m} \sim \text{Exp}(\lambda_{c_m})$, where $\lambda_{1} = \frac{1}{1-\xi}$ and $\lambda_{2} = \frac{1}{1+\xi}$, thus maintaining unitary average rate.
We set the transition matrices to:
\begin{equation}
\bm{\Theta}_1 = \bm{\Theta}_2 = \left[ \begin{matrix}
0.95 & 0.05 \\
0.8 & 0.2
\end{matrix} \right].
\end{equation}
This matrix has a corresponding steady state probability of $\kappa_1 \simeq 0.94$ for state 1 and $\kappa_2 = 1-\kappa_1 \simeq 0.06$ for state 2. We set the exponential parameter for \gls{qs} 1 as before to $\lambda_1 = \frac{1}{1+\xi}$, but in order to maintain the same average capacity we need to set $\lambda_2 = \frac{1 - \kappa_1}{1- (1+\xi) \kappa_1}$. Moreover, in order to maintain positive capacities, it must be $\xi \leq \xi_{max} = \frac{1}{\kappa_1} - 1 = 0.0625$. For this reason, we define the normalized state asymmetry parameter as $\bar{\xi} = \frac{\xi}{\xi_{max}}$. All the other parameters for this scenario are the same that we used in the average load scenario.

\begin{figure*}[h]
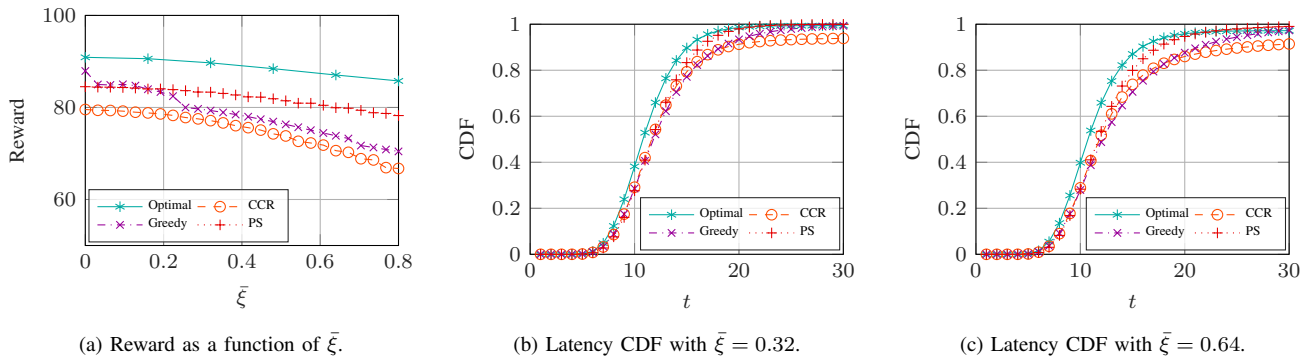

     \centering
          \begin{subfigure}[b]{0.32\textwidth}
         \centering
         \input{tikz_figures/scenario4}
         \caption{Reward as a function of $\bar{\xi}$.}
         \label{fig:markrew}
     \end{subfigure}
     \begin{subfigure}[b]{0.32\textwidth}
         \centering
         \input{tikz_figures/cdf_s4_x032}
         \caption{Latency \gls{cdf} with $\bar{\xi}=0.32$.}
         \label{fig:m_cdf_xi06}
     \end{subfigure}
     \begin{subfigure}[b]{0.32\textwidth}         
         \centering
         \input{tikz_figures/cdf_s4_xi064}
         \caption{Latency \gls{cdf} with $\bar{\xi}=0.64$.}       
         \label{fig:m_cdf_xi03}
     \end{subfigure}
        \caption{Performance in the Markov scenario.}
        \label{fig:markov}
\end{figure*}

The performance we obtained  in this scenario is shown in Fig.~\ref{fig:markov}: if the asymmetry is small, the optimum is still very close to the value of the average load scenario, while other strategies cannot compensate correctly for the variations in the capacity, with a much sharper decline in performance. We can observe that the optimal reward for $\bar{\xi} = 0.8$ is approximately $85\%$, which is close to the percentage of blocks with both \glspl{qs} in state $1$, which is $88.6\%$. This confirms the intuition that for high enough $\bar{\xi}$ the delivery probability depends mostly on the channel state rather than on the statistics of the service time in that state. In fact, if we consider each channel state combination separately, we find out that the success probability given that both \glspl{qs} are in state 1 is $0.94$, and $0.22$ when only one of the channel is in state 2 while with both channels in state 2, in time delivery is basically impossible. We also notice that the heuristic methods are not suffering from quantization effects, as they where doing in scenarios 1, 2 and 3. This is because the channels are in the same state with very high probability ($\kappa_1^2 + \kappa_2^2 \approx 0.9$), so that asymmetry between the channel rates is rare. 
In this case, adding redundancy is not a good policy, as it leads to building up a huge queue when one or both channels are in the low-capacity state, and the \gls{ccr} policy underperforms for this reason.

Interestingly, Fig.~\ref{fig:m_cdf_xi06} shows that the \gls{ps} policy actually has a better latency than the optimal policy if $\tau_d>20$. This is not a violation of the optimality, as the aim of the optimal design is to maximize the probability at $\tau_d = 15$. However, in contrast to what we observed in the previous scenarios, here the choice of the deadline forces the controller to drop some packets, so that the effectiveness of the policy significantly depends on the choice of the deadline.

\begin{figure}[h]
     \centering
          \begin{subfigure}[b]{0.45\textwidth}
         \centering
         \input{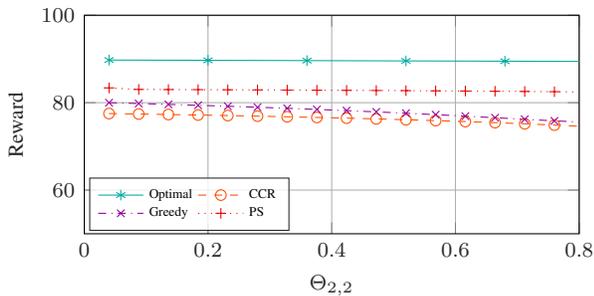}
         \caption{Reward as a function of $\Theta_{2,2}$.}
         \label{fig:sojrew}
     \end{subfigure}
     \begin{subfigure}[b]{0.45\textwidth}
         \centering
         \input{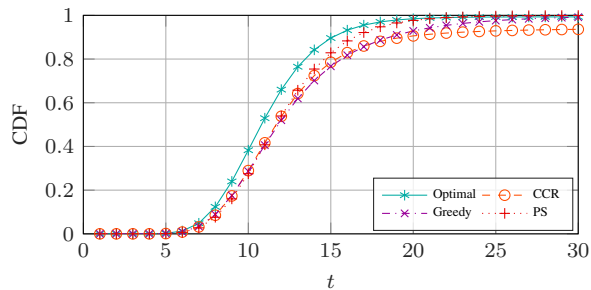}
         \caption{Latency \gls{cdf} with $\Theta_{2,2}=0.36$.}
         \label{fig:m_cdf_th36}
     \end{subfigure}
        \caption{Performance in the Markov scenario with variable sojourn times.}
        \label{fig:sojourn}
\end{figure}

We can also see what happens if we vary the sojourn times of the \glspl{mc}. We fix the steady state probabilities as they were in the Markov scenario and the state asymmetry parameter to $\bar{\xi}=0.32$, then design the transition matrix to change the sojourn time in each state. In particular, given $\kappa_2$ and the transition probability $\Theta_{2,2}$ from state $2$ to state $2$, we compute the transition matrix as:
\begin{equation}
\bm{\Theta}_1 = \bm{\Theta}_2 = \left[
\begin{matrix}
\frac{1 - 2 \kappa_2 - \Theta_{2,2} \kappa_2}{1-\kappa_2} & 1-\Theta_{2,2} \\
1- \frac{1 - 2 \kappa_2 - \Theta_{2,2} \kappa_2}{1-\kappa_2} & \Theta_{2,2}
\end{matrix}
\right].
\end{equation}
It is easy to verify that this matrix has the properties described above.

The results for the variable sojourn time scenario are shown in Fig.~\ref{fig:sojourn}. The reward plot in Fig.~\ref{fig:sojrew} shows that a longer sojourn time on each state of the \gls{mc} affects the optimal and the \gls{ps} strategies only slightly. The \gls{ccr} and greedy strategies seem to be impacted more severely, as they tend to send more redundancy during the long periods with lower capacity. In fact, the more packets are sent in this state, the longer it will take to clear out the backlog when the channel goes back to normal. In particular, this effect highlights the problem of the greedy policy, as it optimizes the chances for the current packet without considering the impact on the future. On the other hand, the optimal policy is barely affected by the length of the sojourn times, and it can outperform the \gls{ps} policy by dropping some blocks during the low-capacity periods.

\subsubsection{Feedback Delay and Error}

Finally, we analyze the effects of channel impairments, substituting the channel with a \gls{pec} or adding a feedback delay $\tau_f$ on both channels. We consider the scenario with average load and add either an error probability or a feedback delay, checking their effect on the reward and, indirectly, on the delivery probability.

\begin{figure}[h]
     \centering
          \begin{subfigure}[b]{0.45\textwidth}
         \centering
         \input{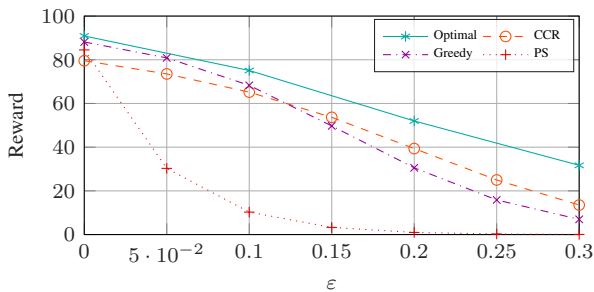}
         \caption{Reward as a function of the channel error probability.}
         \label{fig:err}
     \end{subfigure}
     \begin{subfigure}[b]{0.45\textwidth}
         \centering
         \input{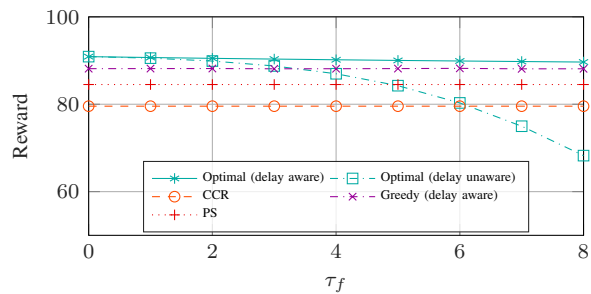}
         \caption{Reward as a function of the feedback delay.}
         \label{fig:del}
     \end{subfigure}
        \caption{Performance in an imperfect channel.}
        \label{fig:impairment}
\end{figure}

Fig.~~\ref{fig:err} shows that all strategies are affected by channel errors, as it requires redundancy just to recover the dropped packets. We can also notice an example of the ``snowball effect'' for the greedy policy: if the error probability is large enough, the greedy policy will increase redundancy and make the queue unstable, reducing the reward because of self-inflicted queuing delay and making the greedy policy worse than \gls{ccr}. Naturally, the \gls{ps} policy performs worse, as it does not include any redundancy and results in a block decoding failure for every dropped packet. As before, the optimal policy significantly outperforms the others, setting the correct amount of redundancy to balance the protection of the current block with the stability of the queue.

Fig.~\ref{fig:del} shows the reward for the different strategies as a function of the feedback delay $\tau_f$. The \gls{ps} and \gls{ccr} strategies, which do not rely on feedback, are unaffected by $\tau_f$. Interestingly, the greedy and optimal strategies are also barely affected if they are aware of the delay, i.e., if the strategies are computed with the correct value of $\tau_f$. However, by using the optimal policy for $\tau_f = 0$ in the case with $\tau_f > 0$, performance quickly becomes even worse than \gls{ps}. 

\subsection{Parameter Sensitivity Analysis}

\begin{figure*}[h]
         \centering
              \begin{subfigure}[b]{0.48\textwidth}
         \centering
         \input{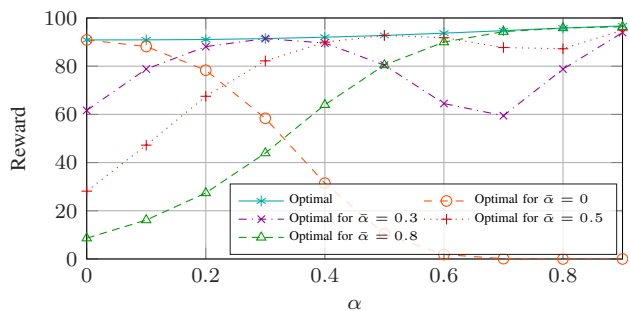}
         \caption{Sensitivity analysis for channel asymmetry.}
         \label{sf:alpha_sens}
     \end{subfigure}
                   \begin{subfigure}[b]{0.48\textwidth}
         \centering
         \input{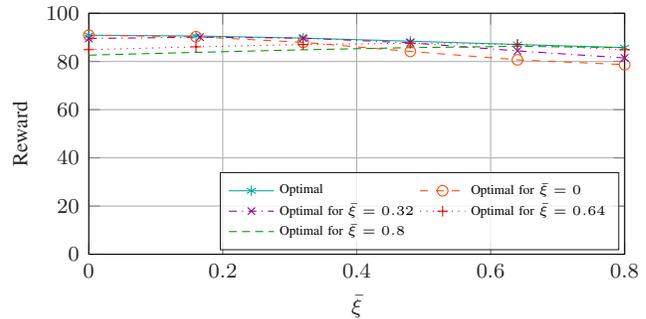}
         \caption{Sensitivity analysis for state asymmetry.}
         \label{sf:xi_sens}
     \end{subfigure}
     
    \begin{subfigure}[b]{0.48\textwidth}
         \centering
         \input{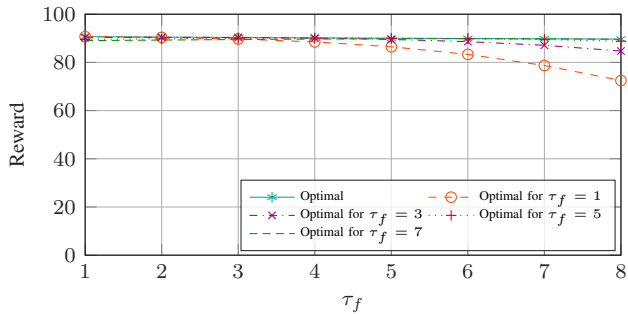}
         \caption{Sensitivity analysis for delayed feedback.}
         \label{sf:tau_sens}
     \end{subfigure}
                   \begin{subfigure}[b]{0.48\textwidth}
         \centering
         \input{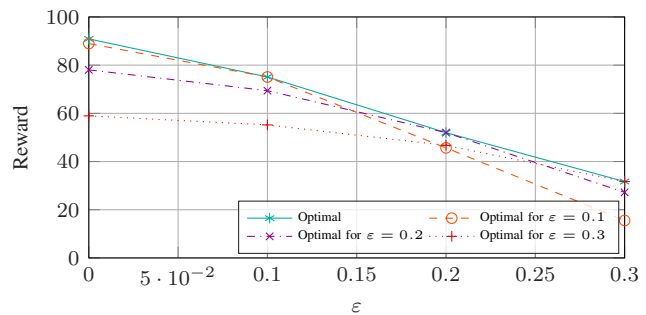}
         \caption{Sensitivity analysis for erasure probability.}
         \label{sf:eps_sens}
     \end{subfigure}
         \caption{Sensitivity analysis for several parameters in the average load scenario.}
         \label{fig:sensasym}
\end{figure*}

We now investigate how much the policies are robust to uncertainty on the knowledge of the \gls{qs} parameters. Fig.~\ref{fig:sensasym} shows what happens if the asymmetry $\alpha$, the Markov state asymmetry $\bar{\xi}$, the feedback delay $\tau_f$, or the erasure probability $\epsilon$ are different from the ones used to compute the policy. Fig.~\ref{sf:alpha_sens} clearly shows that errors in estimating $\alpha$ have the strongest effect, and can significantly impact the reward. It is easy to see how inverting the fast and slow channels might wreak havoc on the strategies in cases with high asymmetry, reducing the reliability significantly. The other parameters are less impacting, although the error rate $\epsilon$ can also have a significant impact, as shown in Fig.~\ref{sf:eps_sens}: this is due to the fact that the error rate alters the required redundancy, as it increases the difference between $\delta_m$ and $\omega_m$. The feedback delay also has an effect on the reward, particularly if it is large: a significant mismatch between the expected and real feedback delays can significantly degrade performance, although not as much as $\varepsilon$ or $\alpha$.

\subsection{Optimal Policy Analysis}\label{ssec:policy_an}

We can also examine in depth the schedules generated by the optimal policy, looking at which states (i.e., queues length at any scheduling time) are visited more often  and how the balance between reliability and low congestion is achieved.

\subsubsection{Load and Capacity Asymmetry} 

First, we analyze the strategies for different values of the asymmetry $\alpha$ with static channels with no error or feedback delay.

\begin{figure*}[t]
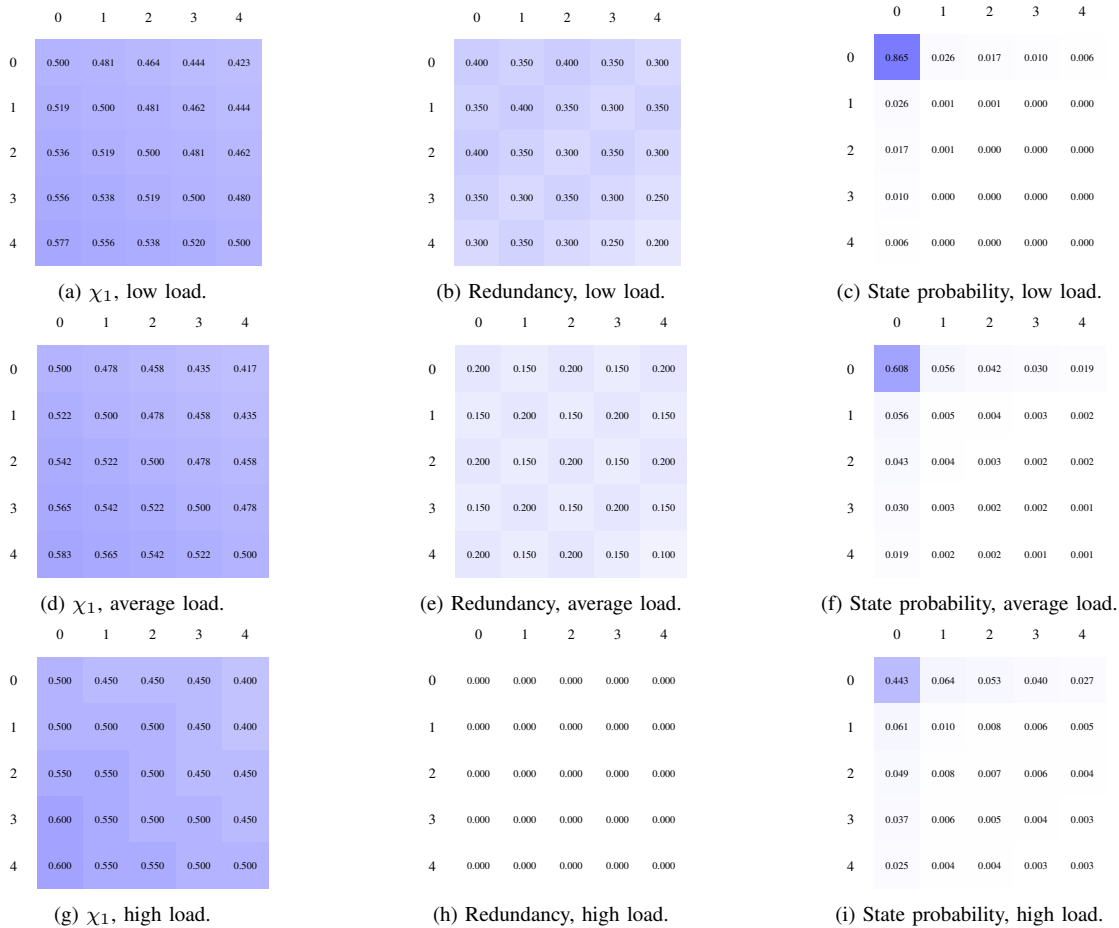

     \centering
     \begin{subfigure}[b]{0.3\textwidth}
         \centering
        \resizebox{0.7\linewidth}{!}{
         \input{tikz_figures/strategies_hmap/channel_asymmetry/fraction_low_x_0_small}
         }
         \caption{$\chi_1$, low load.}
         \label{fig:fractionLowLoad}
     \end{subfigure}
     \begin{subfigure}[b]{0.3\textwidth}         
         \centering
        \resizebox{0.7\linewidth}{!}{
         \input{tikz_figures/strategies_hmap/channel_asymmetry/redundancy_low_x_0_small}
         }
         \caption{Redundancy, low load.}
         \label{fig:redundancyLowLoad}
     \end{subfigure}
     \begin{subfigure}[b]{0.3\textwidth}
         \centering
        \resizebox{0.7\linewidth}{!}{
         \input{tikz_figures/strategies_hmap/channel_asymmetry/probability_low_x_0_small}}
         \caption{State probability, low load.}
         \label{fig:probabilityLowLoad}
     \end{subfigure}
          \begin{subfigure}[b]{0.3\textwidth}
         \centering
        \resizebox{0.7\linewidth}{!}{
         \input{tikz_figures/strategies_hmap/channel_asymmetry/fraction_average_x_0_small}}
         \caption{$\chi_1$, average load.}
         \label{fig:fractionAvLoad}
     \end{subfigure}
     \begin{subfigure}[b]{0.3\textwidth}         
         \centering       
         \resizebox{0.7\linewidth}{!}{
         \input{tikz_figures/strategies_hmap/channel_asymmetry/redundancy_average_x_0_small}}
         \caption{Redundancy, average load.}
         \label{fig:redundancyAvLoad}
     \end{subfigure}
     \begin{subfigure}[b]{0.3\textwidth}
         \centering
         \resizebox{0.7\linewidth}{!}{
         \input{tikz_figures/strategies_hmap/channel_asymmetry/probability_average_x_0_small}}
         \caption{State probability, average load.}
         \label{fig:probabilityAvLoad}
     \end{subfigure}
          \begin{subfigure}[b]{0.3\textwidth}
         \centering      
         \resizebox{0.7\linewidth}{!}{
         \input{tikz_figures/strategies_hmap/channel_asymmetry/fraction_high_x_0_small}}
         \caption{$\chi_1$, high load.}
         \label{fig:fractionHiLoad}
     \end{subfigure}
     \begin{subfigure}[b]{0.3\textwidth}         
         \centering
         \resizebox{0.7\linewidth}{!}{
         \input{tikz_figures/strategies_hmap/channel_asymmetry/redundancy_high_x_0_small}}
         \caption{Redundancy, high load.}
         \label{fig:redundancyHiLoad}
     \end{subfigure}
\begin{subfigure}[b]{0.3\textwidth}
         \centering        
         \resizebox{0.7\linewidth}{!}{
         \input{tikz_figures/strategies_hmap/channel_asymmetry/probability_high_x_0_small}}
         \caption{State probability, high load.}
         \label{fig:probabilityHiLoad}
     \end{subfigure}
        \caption{Strategies for $\alpha = 0$.}
        \label{fig:strategyvsload}
\end{figure*}

Fig.~\ref{fig:strategyvsload} shows three heatmaps representing the fraction of packets the first queue $\chi_1=\frac{s_1}{s_1+s_2}$, the redundancy $N/K$ and the state probability for all states with less than $8$ packets in the queue, in the symmetric case ($\alpha=0$). It is interesting to note that blocks are dropped more often for lower loads, for which the time deadline is looser. This counterintuitive behavior is explained by considering that, if the deadline is tight, dropping a block when the queue is long has marginal effects on the final performance, as that block has a low chance of being delivered anyway. On the other hand, if the deadline is looser, the probability of delivering the block on time is higher. However, these states are very rarely reached in practice, as the right side of the figure shows: while the scenarios with a higher load have a higher probability of reaching longer queues, the scheduling almost always maintains one of the two queues empty, effectively alternating the two \glspl{qs} by placing more packets on the empty queue and reducing redundancy if the queues start filling up. In fact, the optimal policy is to maintain the queues as empty as possible, as any additional redundancy would hurt future blocks by causing self-queuing delay. In this case, and in most of the more complex ones we analyze below, state $(0,0)$ has a very high probability, and the state of the queues changes only in unlucky cases.

In this simple scenario, we can also give some additional results to explain the choices made by the optimal policy. We consider the highest possible reliability that can theoretically be obtained for the next step, given by schedule $\mathbf{s}=(\infty,\infty)$. This action will clearly penalize all future packet blocks, which will find infinitely many packets in the queues. However, it is optimal if we only consider the next step. The upper bound $\delta_m^*$ to the probability of delivering $x$ packets from the current block on path $m$ is then simply given by a shifted Poisson distribution:
\begin{equation}
  \delta_m^*(x|\tau,q_m)=\begin{cases}
                   \frac{\Gamma(q_m+1,\mu_m'\tau)}{q_m!}, &x=0;\\
                   \frac{(\mu_m'\tau)^{q_m+x}e^{-\mu_m'\tau}}{(q_m+x)!}, & x>0;
                  \end{cases}
\end{equation}
where $\mu_m'=\mu_m(1-\varepsilon_m)$, and $\Gamma(m,x)$ is the upper incomplete gamma function. As a block is on time if $K$ or more of its packets arrive by $\tau$, we get the following delivery probability bound:
\begin{equation}
\begin{aligned}
  \rho^*(\tau,\bm{\varepsilon}|\mathbf{q})&=1-\sum_{x_1=0}^{K-1}\sum_{x_2=0}^{K-1-x_1}d_1^*(x_1|\tau,q_1)d_2^*(x_2|\tau,q_2)\\
  &=1-\frac{\Gamma(q_1+q_2+K,(\mu_1'+\mu_2')\tau)}{(q_1+q_2+K-1)!} \\ &-\sum_{j=1}^2
  \frac{\Gamma(q_j+1,\mu_j'\tau)\Gamma(q_{3-j}+K,\mu_{3-j}'\tau)}{q_j!(q_{3-j}+K-1)!}.
\end{aligned}
\end{equation}
This result can simply be achieved by convolving the two Poisson distributions, considering the cases in which one of the two paths delivers no new packets separately. We define the \emph{normalized reliability} $\eta(\mathbf{s},\mathbf{q})$ of schedule $\mathbf{s}$ as $
  \eta(\mathbf{s},\mathbf{q})=\frac{\rho(\tau,\bm{\varepsilon}|\mathbf{q},\mathbf{s})}{\rho^*(\tau,\bm{\varepsilon}|\mathbf{q})}$.
We can also define the \emph{sustainability} of a schedule $\zeta(\mathbf{s})$ as the probability that the queue on each path will decrease or remain stable in the next time step. On a single path, this probability is ì $\zeta_m(s_m)=1-\frac{\Gamma(s_m,\mu_m'\tau)}{(s_m-1)!}$. Since the two paths are independent we can define $\zeta(\mathbf{s})=\zeta_1(s_1)\zeta_2(s_2)$. Naturally, there is a tradeoff between the $\eta$ and $\zeta$ metrics, as schedules with a lower redundancy will tend to be more sustainable, as they require the transmission of fewer packets in time $\tau$ to maintain a stable queue, but their reliability will decrease, as fewer errors or late packets can make the block miss the deadline.

\begin{figure*}[t]
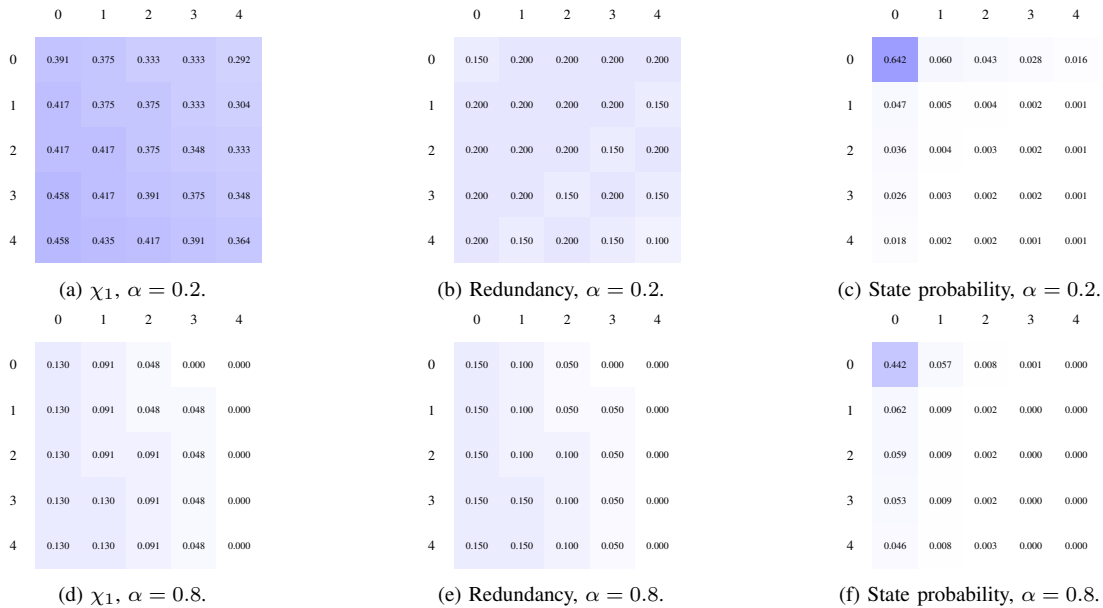

     \centering
     \begin{subfigure}[b]{0.3\textwidth}
         \centering
         \resizebox{0.7\linewidth}{!}{
         \input{tikz_figures/strategies_hmap/channel_asymmetry/fraction_average_x_02_small}
         }
         \caption{$\chi_1$, $\alpha = 0.2$.}
         \label{fig:fractiona2}
     \end{subfigure}
     \begin{subfigure}[b]{0.3\textwidth}         
         \centering
         \resizebox{0.7\linewidth}{!}{
         \input{tikz_figures/strategies_hmap/channel_asymmetry/redundancy_average_x_02_small}
         }
         \caption{Redundancy, $\alpha = 0.2$.}
         \label{fig:redundancya2}
     \end{subfigure}
     \begin{subfigure}[b]{0.3\textwidth}
         \centering
          \resizebox{0.7\linewidth}{!}{
         \input{tikz_figures/strategies_hmap/channel_asymmetry/probability_average_x_02_small}
         }
         \caption{State probability, $\alpha = 0.2$.}
         \label{fig:probabilitya2}
     \end{subfigure}
          \begin{subfigure}[b]{0.3\textwidth}
         \centering
         \resizebox{0.7\linewidth}{!}{
         \input{tikz_figures/strategies_hmap/channel_asymmetry/fraction_average_x_08_small}
         }
         \caption{$\chi_1$, $\alpha = 0.8$.}
         \label{fig:fractiona8}
     \end{subfigure}
     \begin{subfigure}[b]{0.3\textwidth}         
         \centering
         \resizebox{0.7\linewidth}{!}{
         \input{tikz_figures/strategies_hmap/channel_asymmetry/redundancy_average_x_08_small}
         }
         \caption{Redundancy, $\alpha = 0.8$.}
         \label{fig:redundancya8}
     \end{subfigure}
     \begin{subfigure}[b]{0.3\textwidth}
         \centering
          \resizebox{0.7\linewidth}{!}{
         \input{tikz_figures/strategies_hmap/channel_asymmetry/probability_high_x_08_small}
         }
         \caption{State probability, $\alpha = 0.8$.}
         \label{fig:probabilitya8}
     \end{subfigure}
        \caption{Strategies for varying $\alpha$ and average load.}
        \label{fig:strategyvsalpha}
\end{figure*}

\begin{figure}[h]
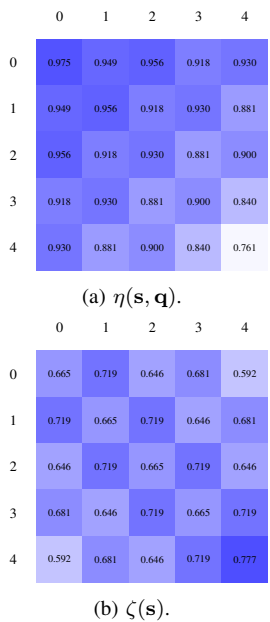

     \centering
     \begin{subfigure}[b]{0.3\textwidth}
         \centering
         \resizebox{0.7\linewidth}{!}{
         \input{tikz_figures/revision/eta.tex}}
         \caption{$\eta(\mathbf{s},\mathbf{q})$.}
         \label{fig:eta}
     \end{subfigure}
     \hspace{0.4cm}
     \begin{subfigure}[b]{0.3\textwidth}         
         \centering
         \resizebox{0.7\linewidth}{!}{
         \input{tikz_figures/revision/zeta.tex}}
         \caption{$\zeta(\mathbf{s})$.}
         \label{fig:zeta}
     \end{subfigure}
        \caption{Optimal tradeoff between reliability and sustainability with $\alpha=0$ in the average load scenario.}
        \label{fig:etazeta}
\end{figure}

The values of $\eta(\mathbf{s},\mathbf{q})$ and $\zeta(\mathbf{s})$ for the optimal policy when $\alpha=0$ in the average load scenario, shown in Fig.~\ref{fig:etazeta}, can help us understand the tradeoff between the reliability of the current block and the sustainability of the schedule. Fig.~\ref{fig:eta} clearly shows that the optimal policy tends to become more conservative as the number of packets in the two queues increases: if the queues are empty, it can add more redundancy without affecting future blocks too heavily. However, as Fig.~\ref{fig:redundancyAvLoad} shows, this does not imply that redundancy is monotonically decreasing. For example, the added redundancy when the queue state is $(1,0)$ is lower than when the state is $(2,0)$. What the normalized reliability shows is that the policy gradually becomes less focused on the current block as the queues increase in size, adding enough redundancy to compensate for the paths' unpredictability but looking more at the future. This is confirmed by Fig.~\ref{fig:zeta}: while normalized reliability uniformly decreases as the queues increase, the sustainability shows an opposite trend. The probability of the queue decreasing or remaining stable tends to grow as the queues become more occupied, taking the system back to a more fruitful state for the next packet. This is not perfectly monotonic, as the optimal policy tends to be more aggressive and concentrate on one path if that path has an empty queue and the other is very full, but the general tradeoff holds, and the choice between immediate and future rewards is highly dependent on the state of the queues. Although this is not shown in the figures, if the queues become extremely big the optimal policy can even decide to drop a block: in this case, it accepts the certainty that the next block will not be delivered, but ``resets'' the queues by letting them empty out and finding a better state at the next block arrival. This case can be relatively frequent in Markov-modulated channels, when some states do not have enough capacity to sustain even the uncoded traffic.

\begin{figure*}[t]
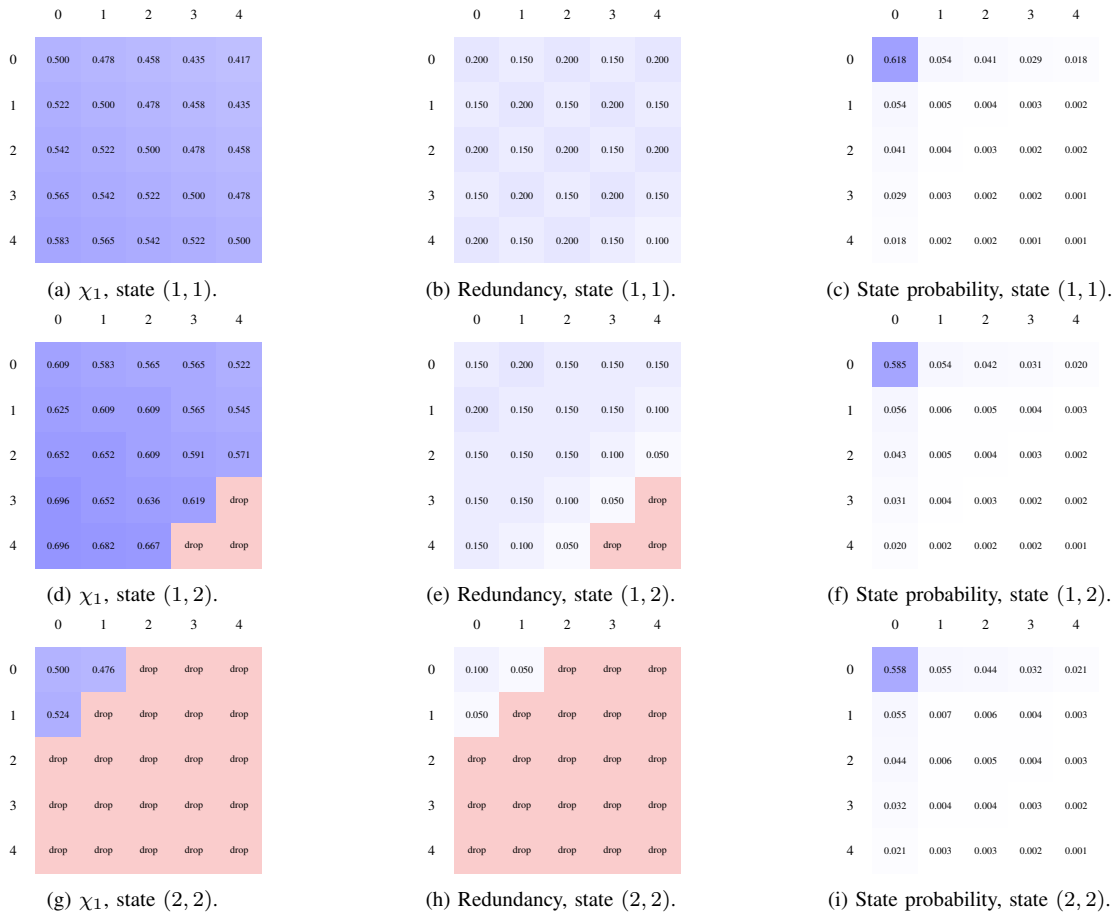

     \centering
     \begin{subfigure}[b]{0.3\textwidth}
         \centering
         \resizebox{0.7\linewidth}{!}{
         \input{tikz_figures/strategies_hmap/stateasymmetry/fraction_stateasymmetry_x_0.32state11_small.tex}
         }
         \caption{$\chi_1$, state $(1,1)$.}
     \end{subfigure}
     \begin{subfigure}[b]{0.3\textwidth}         
         \centering
         \resizebox{0.7\linewidth}{!}{
         \input{tikz_figures/strategies_hmap/stateasymmetry/redundancy_stateasymmetry_x_0.32state11_small.tex}
         }
         \caption{Redundancy, state $(1,1)$.}
     \end{subfigure}
     \begin{subfigure}[b]{0.3\textwidth}
         \centering
          \resizebox{0.7\linewidth}{!}{
         \input{tikz_figures/strategies_hmap/stateasymmetry/probability_stateasymmetry_x_0.32state11_small.tex}
         }
         \caption{State probability, state $(1,1)$.}
     \end{subfigure}
     
          \begin{subfigure}[b]{0.3\textwidth}
         \centering
         \resizebox{0.7\linewidth}{!}{
         \input{tikz_figures/strategies_hmap/stateasymmetry/fraction_stateasymmetry_x_0.32state12_small.tex}
         }
         \caption{$\chi_1$, state $(1,2)$.}
     \end{subfigure}
     \begin{subfigure}[b]{0.3\textwidth}         
         \centering
         \resizebox{0.7\linewidth}{!}{
         \input{tikz_figures/strategies_hmap/stateasymmetry/redundancy_stateasymmetry_x_0.32state12_small.tex}
         }
         \caption{Redundancy, state $(1,2)$.}
     \end{subfigure}
     \begin{subfigure}[b]{0.3\textwidth}
         \centering
          \resizebox{0.7\linewidth}{!}{
         \input{tikz_figures/strategies_hmap/stateasymmetry/probability_stateasymmetry_x_0.32state12_small.tex}
         }
         \caption{State probability, state $(1,2)$.}
     \end{subfigure}
          \begin{subfigure}[b]{0.3\textwidth}
         \centering
         \resizebox{0.7\linewidth}{!}{
         \input{tikz_figures/strategies_hmap/stateasymmetry/fraction_stateasymmetry_x_0.32state22_small.tex}
         }
         \caption{$\chi_1$, state $(2,2)$.}
     \end{subfigure}
     \begin{subfigure}[b]{0.3\textwidth}         
         \centering
         \resizebox{0.7\linewidth}{!}{
         \input{tikz_figures/strategies_hmap/stateasymmetry/redundancy_stateasymmetry_x_0.32state22_small.tex}
         }
         \caption{Redundancy, state $(2,2)$.}
     \end{subfigure}
     \begin{subfigure}[b]{0.3\textwidth}
         \centering
          \resizebox{0.7\linewidth}{!}{
         \input{tikz_figures/strategies_hmap/stateasymmetry/probability_stateasymmetry_x_0.32state22_small.tex}
         }
         \caption{State probability, state $(2,2)$.}
     \end{subfigure}
        \caption{Strategies for the Markov-modulated channels with $\Theta=0.32$.} \vspace{-0.5cm}
        \label{fig:strategyvstate}
\end{figure*}

We next examine the optimal policy in case of asymmetric channels. Fig.~\ref{fig:strategyvsalpha} shows the heatmaps for the average load scenario and two values of $\alpha$, namely, 0.2 and 0.8. It is easy to see that, as $Q_2$ capacity grows, the number of packets on it grows correspondingly, although redundancy decreases: if the first \gls{qs} takes up more and more of the load, and the second one cannot provide additional reliability, it becomes harder to remain in favorable states with short queues, as the heatmaps on the right show. This is similar to a single-path transmission, and reliability is correspondingly lower, as we will see in the sections below.

\subsubsection{Markov-Modulated Capacity} 

In this section, we analyze the schedules obtained with the optimal policy in the scenario described in \ref{sec:markovoptimal}, and with the same parameters for the  heuristic policy used for the average load scenario in section \ref{sec:heursitciasymmetry}.

Fig.~\ref{fig:strategyvstate} shows the same plots we presented above, for the states $(c_1, c_2) \in \{(1,1), (1,2), (2,2)\}$. We omit the state $(2,1)$ as it is symmetric to $(1,2)$. The probability of the queue state probability is conditional on the channel Markov chain state.
Surprisingly, in state $(1,2)$ the threshold for dropping the block and the amount of redundancy only depend on the aggregate number of packets on both queues, whereas the fraction of packets sent to $Q_1$ is the only asymmetric feature of the policy (i.e.,it does not remain the same if we swap the \glspl{qs}).
Moreover, the policy in state $(1,2)$ is significantly less aggressive than that for static asymmetric channels. This yields a higher probability of empty queues in  state ($1,1$). In other words, the optimal policy involves sacrificing some performance in bad states to ensure higher success probabilities in the more favorable states: this is even clearer in state $(2,2)$, as blocks are almost always dropped.

\begin{figure*}[t!]
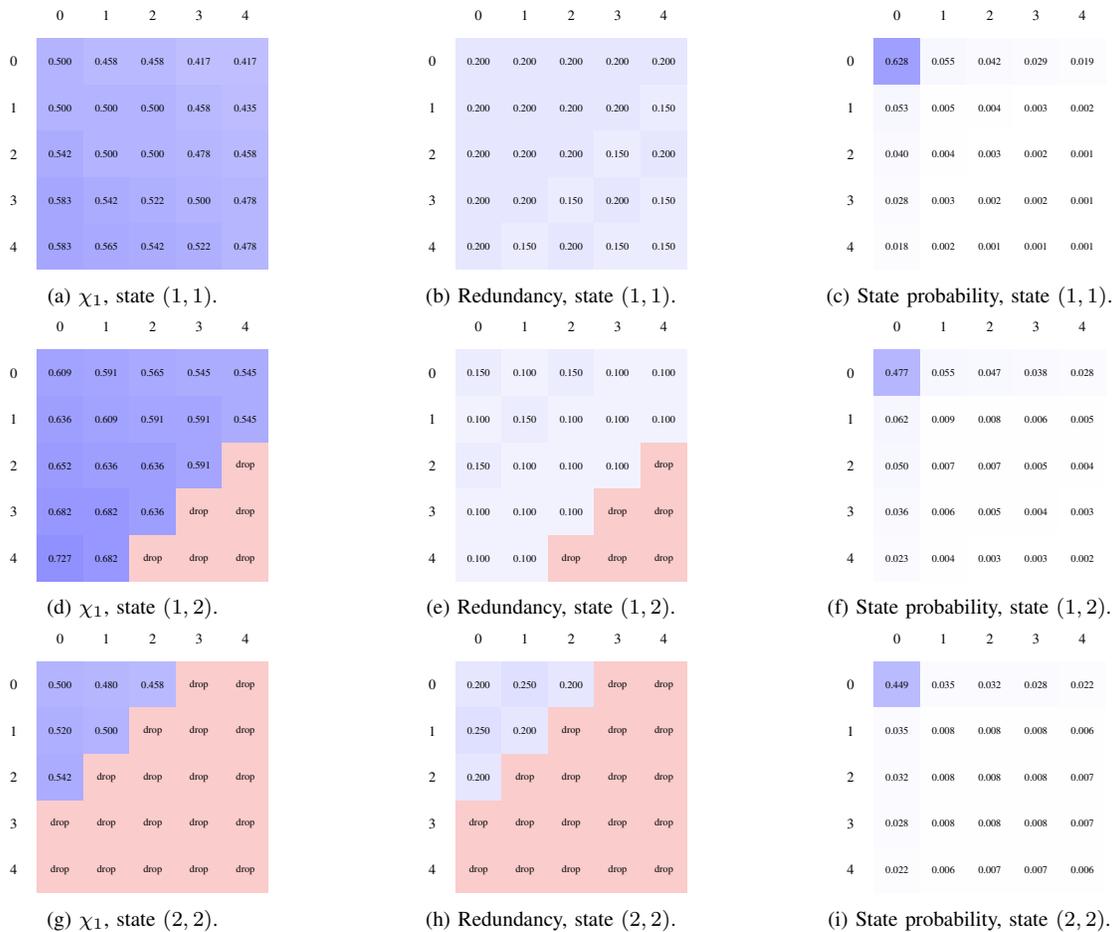

     \centering
     \begin{subfigure}[b]{0.3\textwidth}
         \centering
          \resizebox{0.7\linewidth}{!}{
         \input{tikz_figures/strategies_hmap/sojourntime/fraction_sojourntime_P22_0.84state11_small}}
         \caption{$\chi_1$, state $(1,1)$.}
     \end{subfigure}
     \begin{subfigure}[b]{0.3\textwidth}         
         \centering
          \resizebox{0.7\linewidth}{!}{
         \input{tikz_figures/strategies_hmap/sojourntime/redundancy_sojourntime_P22_0.84state11_small}}
         \caption{Redundancy, state $(1,1)$.}
     \end{subfigure}
     \begin{subfigure}[b]{0.3\textwidth}
         \centering
          \resizebox{0.7\linewidth}{!}{
         \input{tikz_figures/strategies_hmap/sojourntime/probability_sojourntime_P22_0.84state11_small}}
         \caption{State probability, state $(1,1)$.}
     \end{subfigure}
          \begin{subfigure}[b]{0.3\textwidth}
         \centering
          \resizebox{0.7\linewidth}{!}{
         \input{tikz_figures/strategies_hmap/sojourntime/fraction_sojourntime_P22_0.84state12_small}}
         \caption{$\chi_1$, state $(1,2)$.}
     \end{subfigure}
     \begin{subfigure}[b]{0.3\textwidth}         
         \centering
          \resizebox{0.7\linewidth}{!}{
         \input{tikz_figures/strategies_hmap/sojourntime/redundancy_sojourntime_P22_0.84state12_small}}
         \caption{Redundancy, state $(1,2)$.}
     \end{subfigure}
     \begin{subfigure}[b]{0.3\textwidth}
         \centering
          \resizebox{0.7\linewidth}{!}{
         \input{tikz_figures/strategies_hmap/sojourntime/probability_sojourntime_P22_0.84state12_small}}
         \caption{State probability, state $(1,2)$.}
     \end{subfigure}     
     \begin{subfigure}[b]{0.3\textwidth}
         \centering
          \resizebox{0.7\linewidth}{!}{
         \input{tikz_figures/strategies_hmap/sojourntime/fraction_sojourntime_P22_0.84state22_small}}
         \caption{$\chi_1$, state $(2,2)$.}
     \end{subfigure}
     \begin{subfigure}[b]{0.3\textwidth}         
         \centering
          \resizebox{0.7\linewidth}{!}{
         \input{tikz_figures/strategies_hmap/sojourntime/redundancy_sojourntime_P22_0.84state22_small}}
         \caption{Redundancy, state $(2,2)$.}
     \end{subfigure}
     \begin{subfigure}[b]{0.3\textwidth}
         \centering
          \resizebox{0.7\linewidth}{!}{
         \input{tikz_figures/strategies_hmap/sojourntime/probability_sojourntime_P22_0.84state22_small}}
         \caption{State probability, state $(2,2)$.}
     \end{subfigure}
        \caption{Strategies for $\Theta_{2,2} = 0.84$.}
        \label{fig:strategyvstatebigsjtime}
\end{figure*}

If we set $\Theta_{2,2}=0.84$, the sojourn time in state $(2,2)$ significantly increases, but the state is visited much less frequently, so having a bad connection is a rare but long-term event. In this case, blocks are dropped less frequently, and the controller deals with the bad state by putting more packets on the good \gls{qs}. Indeed, as long as the \gls{qs} will remain in the bad state 2, dropping blocks would lead to a very low reliability, and it is better to risk filling up the queue than just waiting for the \gls{qs} to return to a good state. The state probability heatmaps on the right side also show that state $(0,0)$ has a lower probability if there is at least one bad \gls{qs}.

\begin{figure*}[t!]
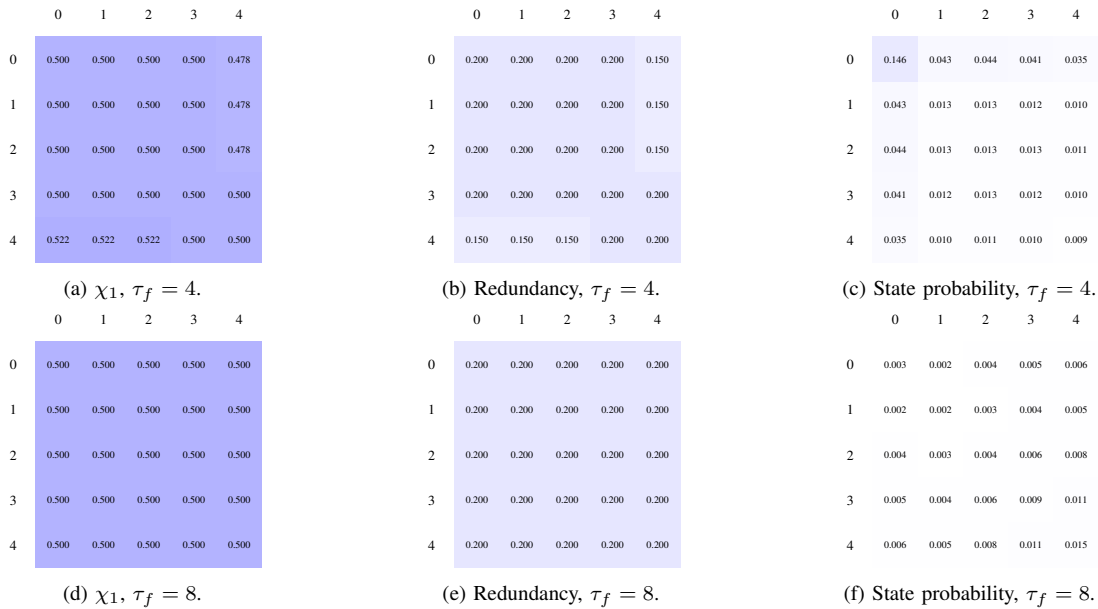

     \centering
          \begin{subfigure}[b]{0.3\textwidth}
         \centering
         \resizebox{0.7\linewidth}{!}{
         \input{tikz_figures/strategies_hmap/delayedfeedback/fraction_delayedfeedback_Tf_4_small}
         }
         \caption{$\chi_1$, $\tau_f = 4$.}
     \end{subfigure}
     \begin{subfigure}[b]{0.3\textwidth}         
         \centering
         \resizebox{0.7\linewidth}{!}{
         \input{tikz_figures/strategies_hmap/delayedfeedback/redundancy_delayedfeedback_Tf_4_small}
         }
         \caption{Redundancy, $\tau_f = 4$.}
     \end{subfigure}
     \begin{subfigure}[b]{0.3\textwidth}
         \centering
          \resizebox{0.7\linewidth}{!}{
         \input{tikz_figures/strategies_hmap/delayedfeedback/probability_delayedfeedback_Tf_4_small}
         }
         \caption{State probability, $\tau_f = 4$.}
     \end{subfigure}
          \begin{subfigure}[b]{0.3\textwidth}
         \centering
         \resizebox{0.7\linewidth}{!}{
         \input{tikz_figures/strategies_hmap/delayedfeedback/fraction_delayedfeedback_Tf_8_small}
         }
         \caption{$\chi_1$, $\tau_f = 8$.}
         \label{subfig:tf8_1}
     \end{subfigure}
     \begin{subfigure}[b]{0.3\textwidth}         
         \centering
         \resizebox{0.7\linewidth}{!}{
         \input{tikz_figures/strategies_hmap/delayedfeedback/redundancy_delayedfeedback_Tf_8_small}
         }
         \caption{Redundancy, $\tau_f = 8$.}
     \end{subfigure}
     \begin{subfigure}[b]{0.3\textwidth}
         \centering
          \resizebox{0.7\linewidth}{!}{
         \input{tikz_figures/strategies_hmap/delayedfeedback/probability_delayedfeedback_Tf_8_small}
         }
         \caption{State probability, $\tau_f = 8$.}
        \label{subfig:tf8_3}
     \end{subfigure}
        \caption{Strategies for varying $\tau_f$.}
        \label{fig:strategyvfeedback}
\end{figure*}

\subsubsection{Delayed Feedback} \label{sec:delayedoptimal}

Finally, we consider the delayed feedback case in the average load scenario. Fig.~\ref{fig:strategyvfeedback} shows the policy for different values of the delay $\tau_f$. The delayed feedback does not have a large effect on the policy, but as $\tau_f$ increases, the state the scheduler sees corresponds to a gradually older picture of the actual state of the queues. If $\tau_f=8$, as in Fig.~\ref{subfig:tf8_1}-\subref{subfig:tf8_3}, the paths can deliver an average of 8 packets before the next block arrives. This leads to the optimal policy accepting longer queues, as the \glspl{qs} have more time to empty them.

\section{Conclusions and Future Work} \label{sec:conc}

In this work, we have presented a model of parallel \glspl{qs} with batch arrivals and latency constraints, deriving the optimal policy in terms of scheduling and packet-level coding to respect the delay constraint over the long term. The tradeoff between adding redundancy to protect the current block and avoiding self-queuing delay is complex, but we show that the optimal policy is robust to parameter estimation errors and significantly outperforms existing heuristic strategies.

The model we present in this work, and the results we presented, show that the decisions on coding and scheduling (i.e., the amount of redundancy needed to protect a block and the paths on which to send the coded packets) are inextricably tied, and are affected by a number of factors in non-trivial ways. The load on the system and the paths' capacities, the erasure probability of each path, the tightness of the deadline, the asymmetry between the available paths, the feedback delay, and the time-varying nature of paths all need to be considered to achieve good performance, even in a relatively simple scenario. In fact, even the simple example presented in Sec.~\ref{sec:analytical} results in a non-trivial set of conditions under which redundancy is beneficial.

The results in more complex cases show that the heuristic strategies often adopted in the literature do not have consistent performance, and even relatively small changes in the scenario might require different settings or even an entirely different approach. In practice, this means that the coding and scheduling policy must be adapted to the specific conditions of the system, e.g., by adapting the amount of redundancy to the load and the tightness of the deadline. Furthermore, a greedy approach that only considers the next block also runs into issues, as it can trigger a \emph{snowball effect} by increasing redundancy more and more as self-queuing delay builds, until the system is completely congested. In some cases, even dropping a block entirely might be warranted to reduce congestion and preserve future performance.

Even though the model does not fully represent a real system (e.g., capacity estimation is assumed to be perfect), it is complex enough to showcase these tradeoffs and for an initial evaluation of practical, foresighted scheduling algorithms.
Future work on this subject might include the translation of the principles and insights from this model into more realistic systems and protocols, as well as applying the optimal policy directly in simpler systems.

\ifCLASSOPTIONcaptionsoff
  \newpage
\fi


\bibliographystyle{IEEEtran}

\bibliography{bibliography.bib}

\appendix

We define a random variable $X$ which can take values in $\mathbb{R}^+$, whose PDF and CDF are denoted as $f(x)$ and $F(x)$, respectively. We then define two values $y$ and $z$, with $y>z>0$. We define the conditioned probability $\xi(z)=P(X\geq y+z|X\geq z)$. The value of $\xi(0)$ is simply given by $\xi(0)=1-F(y)$. We can compute $\xi(z)$ by using Bayes' theorem:
\begin{equation}
\begin{small}
\begin{aligned}
P(X\geq y+z|X\geq z)&=\frac{P(X\geq z|X\geq y+z)P(X\geq y+z)}{P(X\geq z)} \\ &=\frac{1-F(y+z)}{1-F(z)}.
\end{aligned}
\end{small}
\end{equation}
We can then give the condition for which $\xi(z)$ is monotonically decreasing:
\begin{equation}
    \frac{f(y+z)}{1-F(y+z)}-\frac{f(z)}{1-F(z)}\geq 0.
\end{equation}
This is a condition on the hazard rate function $\nu(z)=\frac{f(z)}{1-F(z)}$, often used in reliability applications~\cite{barlow1963properties}.
If $\nu(z)$ is monotonically non-decreasing, $\xi(z)$ is monotonically non-increasing, i.e., the conditional probability of $X$ being larger than $y+z$ if we already know that it is larger than $z$ decreases as $z$ increases. The hazard rate is often hard to compute, but it is monotonically increasing for normal distributions, Gamma and Weibull distributions with a shape parameter $\alpha>1$. The exponential distribution has a constant hazard rate, as it is memoryless. A decreasing hazard rate is sometimes used as the definition of heavy-tailedness.

\end{document}